%% file: main.tex
\documentclass[10pt,twocolumn,letterpaper]{article}

\usepackage[pagenumbers]{cvpr} % To force page numbers, e.g. for an arXiv version
\usepackage[dvipsnames]{xcolor}

\definecolor{cvprblue}{rgb}{0.21,0.49,0.74}
\usepackage[pagebackref,breaklinks,colorlinks,citecolor=cvprblue]{hyperref}

\usepackage{graphicx}
\usepackage{amsmath}
\usepackage{amssymb}
\usepackage{booktabs}
\usepackage{multicol}
\usepackage{multirow}
\usepackage{comment}
\usepackage{float}
\usepackage[toc,page,header]{appendix}
\usepackage{minitoc}
\usepackage{etoc}
\usepackage{appendix}
\usepackage{etoc}
\usepackage{gensymb}
\usepackage[accsupp]{axessibility} % Improves PDF readability for those with visual impairments.

\usepackage[capitalize]{cleveref}
\crefname{section}{Sec.}{Secs.}
\Crefname{section}{Section}{Sections}

\Crefname{table}{Table}{Tables}
\crefname{table}{Tab.}{Tabs.}

\newcommand{\name}{\textsc{DiffRIR}\xspace}

\definecolor{deeppurple}{rgb}{0.1, 0.0, 0.45}

\definecolor{deepgreen}{rgb}{0.0, 0.75, 0.96}

\definecolor{nicegreen}{rgb}{0.1, 0.6, 0.2}

\definecolor{navy}{rgb}{0.0, 0.0, 0.5}
\definecolor{neongreen}{rgb}{0.22, 0.88, 0.08}

\newcommand{\cameraready}[1]{{\color{black}#1}}
\newcommand{\camerareadysection}[1]{{\color{black}#1}}
\newcommand{\myparagraph}[1]{\vspace{0pt}\paragraph{#1}}

\makeatletter
\renewcommand{\paragraph}{%
\@startsection{paragraph}{4}%
{\z@}{0.4em}{-1em}%
{\normalfont\normalsize\bfseries}%
}
\makeatother

\newcommand\rurl[1]{%
  \href{https://#1}{\nolinkurl{#1}}%
}

\title{Hearing Anything Anywhere}

\author{
Mason Long Wang$^1$\thanks{Equal contribution.}
\quad
Ryosuke Sawata$^{1,2}$$^*$
\quad
Samuel Clarke$^1$
\\
Ruohan Gao$^{1,3}$
\quad
Shangzhe Wu$^1$
\quad
Jiajun Wu$^1$ \\[0.3em]
$^1$Stanford University
\quad
$^2$Sony AI
\quad
$^3$University of Maryland, College Park
\\
{\small\rurl{masonlwang.com/hearinganythinganywhere}}
}

\begin{document}
\maketitle

\input{shortcut}

\input{0_abstract}
\input{1_intro}
\input{2_related}
\input{4_method}
\input{3_dataset}

\input{5_experiments}
\input{6_conclusions}

% \small
% \myparagraph{Acknowledgments.}
\paragraph{Acknowledgments.}
The work is in part supported by NSF CCRI \#2120095, RI \#2211258, RI \#2338203, ONR MURI N00014-22-1-2740, Adobe, Amazon, and Sony.

% REFERENCES
{
    \small
    \bibliographystyle{ieeenat_fullname}
    \bibliography{ref}
}

\clearpage

\twocolumn[{%
  \centering
  \Large{\textbf{Hearing Anything Anywhere}} \\
  \vspace{0.5em}
  \large{\textbf{-- Supplementary Material --}} \\
  \vspace{0.5em}
  {\normalsize{\textbf{Website: }} \small{\rurl{masonlwang.com/hearinganythinganywhere}}}
  \vspace{0.5em}
  
  {\normalsize{\textbf{Code: }}\small{\rurl{github.com/maswang32/hearinganythinganywhere}}}
  \vspace{0.5em}
  
  {\normalsize{\textbf{Dataset: }}\small{\rurl{zenodo.org/records/11195833}}}
  \vspace{0.5em}

  {\normalsize{\textbf{Video: }}\small{\rurl{youtube.com/watch?v=Cv9oOFVXem4}}}

  \vspace{2em}
}]

{
  \hypersetup{linkcolor=black}
  \tableofcontents
}

\input{7_appendix}

\end{document}

%% file: shortcut.tex
\newcommand{\netname}{\textbf{RICO}}

\newcommand{\ba}{\mathbf{a}}\newcommand{\bA}{\mathbf{A}}
\newcommand{\bb}{\mathbf{b}}\newcommand{\bB}{\mathbf{B}}
\newcommand{\bc}{\mathbf{c}}\newcommand{\bC}{\mathbf{C}}
\newcommand{\bd}{\mathbf{d}}\newcommand{\bD}{\mathbf{D}}
\newcommand{\be}{\mathbf{e}}\newcommand{\bE}{\mathbf{E}}
\newcommand{\bff}{\mathbf{f}}\newcommand{\bF}{\mathbf{F}} %
\newcommand{\bg}{\mathbf{g}}\newcommand{\bG}{\mathbf{G}}
\newcommand{\bh}{\mathbf{h}}\newcommand{\bH}{\mathbf{H}}
\newcommand{\bi}{\mathbf{i}}\newcommand{\bI}{\mathbf{I}}
\newcommand{\bj}{\mathbf{j}}\newcommand{\bJ}{\mathbf{J}}
\newcommand{\bk}{\mathbf{k}}\newcommand{\bK}{\mathbf{K}}
\newcommand{\bl}{\mathbf{l}}\newcommand{\bL}{\mathbf{L}}
\newcommand{\bm}{\mathbf{m}}\newcommand{\bM}{\mathbf{M}}
\newcommand{\bn}{\mathbf{n}}\newcommand{\bN}{\mathbf{N}}
\newcommand{\bo}{\mathbf{o}}\newcommand{\bO}{\mathbf{O}}
\newcommand{\bp}{\mathbf{p}}\newcommand{\bP}{\mathbf{P}}
\newcommand{\bq}{\mathbf{q}}\newcommand{\bQ}{\mathbf{Q}}
\newcommand{\br}{\mathbf{r}}\newcommand{\bR}{\mathbf{R}}
\newcommand{\bs}{\mathbf{s}}\newcommand{\bS}{\mathbf{S}}
\newcommand{\bt}{\mathbf{t}}\newcommand{\bT}{\mathbf{T}}
\newcommand{\bu}{\mathbf{u}}\newcommand{\bU}{\mathbf{U}}
\newcommand{\bv}{\mathbf{v}}\newcommand{\bV}{\mathbf{V}}
\newcommand{\bw}{\mathbf{w}}\newcommand{\bW}{\mathbf{W}}
\newcommand{\bx}{\mathbf{x}}\newcommand{\bX}{\mathbf{X}}
\newcommand{\by}{\mathbf{y}}\newcommand{\bY}{\mathbf{Y}}
\newcommand{\bz}{\mathbf{z}}\newcommand{\bZ}{\mathbf{Z}}

\newcommand{\balpha}{\boldsymbol{\alpha}}\newcommand{\bAlpha}{\boldsymbol{\Alpha}}
\newcommand{\bbeta}{\boldsymbol{\beta}}\newcommand{\bBeta}{\boldsymbol{\Beta}}
\newcommand{\bgamma}{\boldsymbol{\gamma}}\newcommand{\bGamma}{\boldsymbol{\Gamma}}
\newcommand{\bdelta}{\boldsymbol{\delta}}\newcommand{\bDelta}{\boldsymbol{\Delta}}
\newcommand{\bepsilon}{\boldsymbol{\epsilon}}\newcommand{\bEpsilon}{\boldsymbol{\Epsilon}}
\newcommand{\bzeta}{\boldsymbol{\zeta}}\newcommand{\bZeta}{\boldsymbol{\Zeta}}
\newcommand{\beeta}{\boldsymbol{\eta}}\newcommand{\bEta}{\boldsymbol{\Eta}} %
\newcommand{\btheta}{\boldsymbol{\theta}}\newcommand{\bTheta}{\boldsymbol{\Theta}}
\newcommand{\biota}{\boldsymbol{\iota}}\newcommand{\bIota}{\boldsymbol{\Iota}}
\newcommand{\bkappa}{\boldsymbol{\kappa}}\newcommand{\bKappa}{\boldsymbol{\Kappa}}
\newcommand{\blambda}{\boldsymbol{\lambda}}\newcommand{\bLambda}{\boldsymbol{\Lambda}}
\newcommand{\bmu}{\boldsymbol{\mu}}\newcommand{\bMu}{\boldsymbol{\Mu}}
\newcommand{\bnu}{\boldsymbol{\nu}}\newcommand{\bNu}{\boldsymbol{\Nu}}
\newcommand{\bxi}{\boldsymbol{\xi}}\newcommand{\bXi}{\boldsymbol{\Xi}}
\newcommand{\bomikron}{\boldsymbol{\omikron}}\newcommand{\bOmikron}{\boldsymbol{\Omikron}}
\newcommand{\bpi}{\boldsymbol{\pi}}\newcommand{\bPi}{\boldsymbol{\Pi}}
\newcommand{\brho}{\boldsymbol{\rho}}\newcommand{\bRho}{\boldsymbol{\Rho}}
\newcommand{\bsigma}{\boldsymbol{\sigma}}\newcommand{\bSigma}{\boldsymbol{\Sigma}}
\newcommand{\btau}{\boldsymbol{\tau}}\newcommand{\bTau}{\boldsymbol{\Tau}}
\newcommand{\bypsilon}{\boldsymbol{\ypsilon}}\newcommand{\bYpsilon}{\boldsymbol{\Ypsilon}}
\newcommand{\bphi}{\boldsymbol{\phi}}\newcommand{\bPhi}{\boldsymbol{\Phi}}
\newcommand{\bchi}{\boldsymbol{\chi}}\newcommand{\bChi}{\boldsymbol{\Chi}}
\newcommand{\bpsi}{\boldsymbol{\psi}}\newcommand{\bPsi}{\boldsymbol{\Psi}}
\newcommand{\bomega}{\boldsymbol{\omega}}\newcommand{\bOmega}{\boldsymbol{\Omega}}

\newcommand{\nA}{\mathbb{A}}
\newcommand{\nB}{\mathbb{B}}
\newcommand{\nC}{\mathbb{C}}
\newcommand{\nD}{\mathbb{D}}
\newcommand{\nE}{\mathbb{E}}
\newcommand{\nF}{\mathbb{F}}
\newcommand{\nG}{\mathbb{G}}
\newcommand{\nH}{\mathbb{H}}
\newcommand{\nI}{\mathbb{I}}
\newcommand{\nJ}{\mathbb{J}}
\newcommand{\nK}{\mathbb{K}}
\newcommand{\nL}{\mathbb{L}}
\newcommand{\nM}{\mathbb{M}}
\newcommand{\nN}{\mathbb{N}}
\newcommand{\nO}{\mathbb{O}}
\newcommand{\nP}{\mathbb{P}}
\newcommand{\nQ}{\mathbb{Q}}
\newcommand{\nR}{\mathbb{R}}
\newcommand{\nS}{\mathbb{S}}
\newcommand{\nT}{\mathbb{T}}
\newcommand{\nU}{\mathbb{U}}
\newcommand{\nV}{\mathbb{V}}
\newcommand{\nW}{\mathbb{W}}
\newcommand{\nX}{\mathbb{X}}
\newcommand{\nY}{\mathbb{Y}}
\newcommand{\nZ}{\mathbb{Z}}

\newcommand{\cA}{\mathcal{A}}
\newcommand{\cB}{\mathcal{B}}
\newcommand{\cC}{\mathcal{C}}
\newcommand{\cD}{\mathcal{D}}
\newcommand{\cE}{\mathcal{E}}
\newcommand{\cF}{\mathcal{F}}
\newcommand{\cG}{\mathcal{G}}
\newcommand{\cH}{\mathcal{H}}
\newcommand{\cI}{\mathcal{I}}
\newcommand{\cJ}{\mathcal{J}}
\newcommand{\cK}{\mathcal{K}}
\newcommand{\cL}{\mathcal{L}}
\newcommand{\cM}{\mathcal{M}}
\newcommand{\cN}{\mathcal{N}}
\newcommand{\cO}{\mathcal{O}}
\newcommand{\cP}{\mathcal{P}}
\newcommand{\cQ}{\mathcal{Q}}
\newcommand{\cR}{\mathcal{R}}
\newcommand{\cS}{\mathcal{S}}
\newcommand{\cT}{\mathcal{T}}
\newcommand{\cU}{\mathcal{U}}
\newcommand{\cV}{\mathcal{V}}
\newcommand{\cW}{\mathcal{W}}
\newcommand{\cX}{\mathcal{X}}
\newcommand{\cY}{\mathcal{Y}}
\newcommand{\cZ}{\mathcal{Z}}

\newcommand{\figref}[1]{Fig.~\ref{#1}}
\newcommand{\secref}[1]{Section~\ref{#1}}
\newcommand{\algref}[1]{Algorithm~\ref{#1}}
\newcommand{\eqnref}[1]{Eq.~\eqref{#1}}
\newcommand{\tabref}[1]{Table~\ref{#1}}

\def\mc{\mathcal}
\def\mb{\mathbf}

\newcommand{\T}{^{\raisemath{-1pt}{\mathsf{T}}}}

\newcommand{\Perp}{\perp\!\!\! \perp}

\makeatletter
\DeclareRobustCommand\onedot{\futurelet\@let@token\@onedot}
\def\@onedot{\ifx\@let@token.\else.\null\fi\xspace}
\def\eg{e.g.,\xspace} \def\Eg{E.g.,\xspace}
\def\ie{i.e.,\xspace} \def\Ie{I.e.,\xspace}
\def\cf{cf\onedot} \def\Cf{Cf\onedot}
\def\etc{etc\onedot}
\def\vs{vs\onedot}
\def\wrt{wrt\onedot}
\def\dof{d.o.f\onedot}
\def\etal{et~al\onedot}
\def\iid{i.i.d\onedot}
\def\evs{\emph{vs}\onedot}
\makeatother

\newcommand*\rot{\rotatebox{90}}

\newcommand{\boldparagraph}[1]{\vspace{0.4em}\noindent{\bf #1:}}

\definecolor{darkgreen}{rgb}{0,0.7,0}
\definecolor{lightred}{rgb}{1.,0.5,0.5}

%% file: 0_abstract.tex
\begin{abstract}

    Recent years have seen immense progress in 3D computer vision and computer graphics, with emerging tools that can virtualize real-world 3D environments for numerous Mixed Reality (XR) applications. However, alongside immersive visual experiences, immersive auditory experiences are equally vital to our holistic perception of an environment. In this paper, we aim to reconstruct the spatial acoustic characteristics of an arbitrary environment given only a sparse set of (roughly 12) room impulse response (RIR) recordings and a planar reconstruction of the scene, a setup that is easily achievable by ordinary users. To this end, we introduce \name, a differentiable RIR rendering framework with interpretable parametric models of salient acoustic features of the scene, including sound source directivity and surface reflectivity. This allows us to synthesize novel auditory experiences through the space with any source audio. To evaluate our method, we collect a dataset of RIR recordings \cameraready{and music} in four diverse, real environments. We show that our model outperforms state-of-the-art baselines on rendering monaural and binaural RIRs and music at unseen locations, and learns physically interpretable parameters characterizing acoustic properties of the sound source and surfaces in the scene. %\samuel{Should we put a concluding sentence that drives home its usefulness for the motivating task?}

\end{abstract}

%% file: 1_intro.tex
\addtocontents{toc}{\protect\setcounter{tocdepth}{0}}
\section{Introduction}
\addtocontents{toc}{\protect\setcounter{tocdepth}{2}}

Much of the impetus to realize immersive virtual reality (VR) stems from the desire to recreate and share \emph{real} scenes and experiences.
Motivated by this goal, recent progress in 3D computer vision and computer graphics has led to tools that can virtualize real-world 3D environments using simple consumer devices (\eg cellphone cameras) for numerous Mixed Reality (XR) applications. Alongside immersive visual experiences, immersive auditory experiences are equally vital to our holistic perception of an environment.
For instance, while the interior of Carnegie Hall in New York City is visually beautiful, one cannot fully appreciate the majesty of its design without experiencing a musical performance in-person and hearing its unique acoustics.

In this paper, our goal is to capture the acoustic intrinsics of a real-world scene using a sparse set of measurements, in order to render arbitrary source audio at any location, hence the name, ``Hearing Anything Anywhere". This is analogous to the task of sparse-view novel view synthesis (NVS) in computer vision and graphics~\cite{mildenhall2021nerf, avidan1997novel,zhou2016view}.

However, there are two key differences between light and sound that make common approaches to visual NVS inapplicable to audio.
First,
light is typically emitted from continuous sources and travels steadily and almost instantly through space, resulting in a largely stationary visual scene. 
In contrast, sound signals are usually time-varying and travel through space at a much slower pace, resulting in a constantly changing 4D acoustic field with both numerous early reflections and late reverberations.
Second, a single camera captures \emph{millions} of pixels in a split second, each recording a distinct light ray from a \emph{particular} direction.
In contrast, a typical microphone only records an amalgamation of sound waves arriving to a \emph{single} location from \emph{all} directions, with different times-of-arrival.
Therefore, while it is possible to capture the appearance of a 3D scene by simply walking through it with a camera, the same approach falls short to record the entire 4D acoustic field.

Thus, capturing a fully immersive acoustic field often necessitates setting up hundreds of microphones densely across the space~\cite{richard2022deep, ratnarajah21_interspeech, luo2022learning, su2022inras}, which is impractical for many consumer use cases. In this work, we attempt to capture real-world acoustic spaces with a \cameraready{\emph{basic}} hardware setup, \eg 12 microphones, which can be easily scaled to arbitrary environments.

To capture the acoustic properties of the scene, we measure a room impulse response (RIR) between the sound source and each microphone location. An RIR is a time-series signal that estimates how a perfect impulse emitted from the source, traveling and bouncing in the room, would be perceived at the listener location. RIRs effectively capture a room's intrinsic acoustic properties between source and listener points, and are thus widely used in acoustic simulation~\citep{RIRjust}. In order to simulate the sound of an arbitrary source for a \cameraready{particular} listener location in a room, the RIR \cameraready{associated with} the source-listener pair is simply convolved with the source audio~\citep{ConvolveRIRs}.

We thus formulate our \emph{Hearing Anything Anywhere} task as inferring RIRs \cameraready{and music} at novel listener locations from a \cameraready{sparse set} of RIRs measured between a single source and a small set of microphone locations spatially distributed within the scene.
Towards this goal, we introduce a fully differentiable impulse response rendering framework \name that reasons about the individual contributions of each acoustic reflection path between the source and the receiver, including the time delay and magnitude of the sound on each path, as well as the influence of reflections from each surface in the scene.

By explicitly modeling the sound source location, the directivity map of the source, and the reflection properties of the surfaces in the scene in a fully differentiable audio rendering framework, we can characterize the parameters of each model through an analysis-by-synthesis paradigm by optimizing the output of \name against the known subset of measured RIRs. After optimizing the interpretable parameters of our model, we can estimate the RIR from any unseen location in the scene.

To validate our method, we collect a dataset that contains RIR measurements from four real-world environments \cameraready{that represent a diverse range of room materials, shape, and complexity}.
Through experiments comparing our framework with current state-of-the-art methods, \name shows greater robustness in real, data-limited scenarios. Moreover, with the explicit and interpretable models of source and surface reflection properties, we can easily synthesize novel auditory experiences with different speaker orientations and locations, which can be useful in applications such as virtual reality and acoustics-aware interior design. In addition, the differentiable and interpretable models of our framework allow us to estimate acoustic parameters of the sound source and surfaces in the room, which can be useful in applications like robotics and architectural design for acoustics.

Our contributions are threefold. First, we contribute \name, a differentiable acoustic inverse rendering framework that can recover the fully immersive acoustic field of a room from a set of 12 sparsely located RIR measurements. Second, we contribute a new dataset of real-world RIRs measured from hundreds of locations in four different real environments. Third, we compare our method to existing methods across various settings, demonstrating that our method is more effective than existing methods on real data in our data-limited scenarios, predicting more accurate RIRs \cameraready{and music} at unseen locations. Code and data are available at the \href{https://masonlwang.com/hearinganythinganywhere}{project website}.

%% file: 2_related.tex
\addtocontents{toc}{\protect\setcounter{tocdepth}{0}}
\section{Related Work}
\addtocontents{toc}{\protect\setcounter{tocdepth}{2}}
\myparagraph{Learning-Based Room Acoustics Prediction.}
While many acoustical learning frameworks model room acoustics implicitly, others explicitly interpolate and predict RIRs at novel points. Frameworks that predict RIRs at novel points in a room vary not only in their underlying techniques, but also in their inputs. 
Some methods do not use vision or geometry to make their estimates, but instead learn to directly approximate a function mapping spatial coordinates to RIRs~\cite{richard2022deep, ratnarajah21_interspeech}.
These methods can require large training set sizes on the order of 1,000 RIRs from a room to effectively interpolate RIRs to novel points within the same room. Alternatively, some methods use geometric features of the scene~\cite{luo2022learning}, such as~\cite{su2022inras}, which learns a diffuse reflection model from a small subset of points in the mesh of the environment, to achieve a performance improvement over pure audio-based methods. Our method uses environment geometry to explicitly model specular reflections on each surface. To validate our approach, we compare against three baselines, including one audio-only method~\cite{richard2022deep} and two methods that use scene geometry~\cite{luo2022learning, su2022inras}.

\myparagraph{Audio-Visual (AV) Room Acoustics Prediction.}
Other methods learn relationships between visual inputs and room acoustics to perform tasks such as predicting the dereverberated signal from an audio recording and a panoramic image of the recording environment~\cite{chen2023learning}, or predicting how an input audio signal would sound in a target space based on an image of the space~\cite{chen2022visual}.
Many works use visual inputs to explicitly perform the novel view acoustic synthesis (NVAS) task. 
For instance, \citet{chen2023novel} proposed the Visually-Guided Acoustic Synthesis (ViGAS) network, which outputs the spatial audio of the speech of a human in corresponding visual frames.
Furthermore, by using audio-visual features as well as geometric ones, \citet{ahn2023novelview} show that the important sub-tasks of NVAS, \eg sound source localization, separation, and dereverberation, can be jointly solved.
AV-NeRF~\cite{liang2023avnerf} improved the performance of both NVS and NVAS tasks via multi-task training by using an audio-based Neural Radiance Field (NeRF). Their audio NeRF estimates variations in the magnitudes of audio perceived from varying locations, whereas we explicitly estimate the RIR, a much more holistic characterization of the environment acoustic properties.

Similar to our binaural prediction task, \citet{garg2023visually} predict binaural audio from an AV scene's monaural audio and visual features extracted from the scene's video frames.
Although AV approaches can sometimes outperform uni-modal audio-only models at estimating environment acoustics, collecting large enough datasets of synchronized audio-visual pairs for these models can be laborious. Perhaps for this reason, many such models, even one boasting few-shot generalization~\cite{majumder2022few}, present results from evaluating exclusively on simulated data.

\myparagraph{Geometry-Based RIR Simulation.}
Many of the aforementioned works use datasets of simulated RIRs generated by the SoundSpaces framework~\cite{chen22soundspaces2}, a fast acoustic simulator based on geometric acoustic methods. They simulate the acoustics of virtualized versions of real rooms from datasets of meshes reconstructed from RGBD scans of real rooms in home and workplace environments, such as the Matterport3D dataset~\cite{chang2017matterport3d} or the Replica dataset~\cite{straub2019replica}. The Geometric-Wave Acoustic (GWA) dataset uses a hybrid propagation algorithm combining wave-based methods~\cite{hamilton2021pffdtd} 
 with geometric acoustic methods, intending to model low-frequency wave effects more accurately, albeit at the cost of longer run-time. The input meshes are from a dataset of professionally designed virtual home layouts \cite{fu20213d}. The Mesh2IR framework uses the GWA dataset to learn a conditional generative adversarial network (cGAN) to more quickly predict RIRs from meshes of rooms~\cite{ratnarajah2022mesh2ir}. The authors do not show how their cGAN's estimates of RIRs compare to measured RIRs from real rooms.

%  \samuel{Look into putting Odeon simulator in here.}

\myparagraph{Differentiable Acoustics.}
The previously mentioned simulators are not differentiable, which precludes gradient-based optimization techniques which can be used in solving inverse problems. Differentiable audio rendering techniques have been used to solve such inverse problems estimating acoustic properties of musical instruments~\cite{engel2020ddsp} and everyday objects~\cite{clarke2022diffimpact}, as well as the reverberation properties of the environments they are in. The authors of~\cite{chitre2023differentiable} implemented a differentiable acoustic ray tracer for inverse tasks in underwater acoustics, such as estimating the absorption of the seabed on simulated 2D data. We use similar principles for estimating absorption parameters of surfaces in 3D environments from our real, airborne sound data.

%% file: 4_method.tex
\begin{figure*}
    \centering
    \includegraphics[trim={60px 20px 130px 10px}, clip, width=0.9\linewidth]{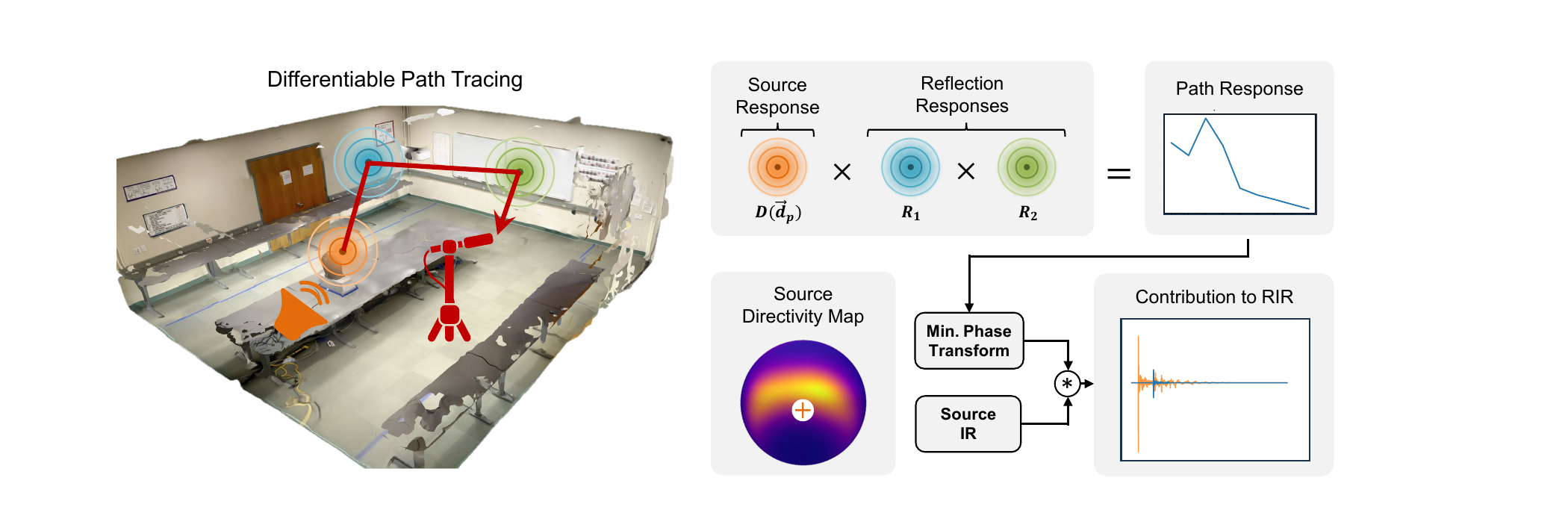}
    \vspace{-5pt}
    \caption{
    \textbf{Differentiable Room Impulse Response Rendering Framework (\name).}
    Our model renders the contribution to the RIR of a single traced reflection path.
    After computing a reflection path, we characterize it by the direction at which it exits the speaker, its length, and the surfaces on which it reflects. The sound source has a learned frequency response that depends on the outgoing direction, and each surface has a different learned frequency response. We multiply each of these responses to estimate the overall path response. To determine the reflection path's time-domain contribution to the final RIR, we apply a minimum-phase inverse-Fourier transform to the path response, convolve it with the source impulse response, and then shift the result in time based on the path length and the speed of sound.}
    \vspace{-5pt}
    \label{fig:DiffRIRFigure}
\end{figure*}

\addtocontents{toc}{\protect\setcounter{tocdepth}{0}}
\section{Method}
\addtocontents{toc}{\protect\setcounter{tocdepth}{2}}
We first lay out the definition of our task, and then introduce our proposed \name framework to approach it.

\addtocontents{toc}{\protect\setcounter{tocdepth}{1}}
\subsection{Task Formulation}\label{sec:background}
\addtocontents{toc}{\protect\setcounter{tocdepth}{2}}

To achieve our goal of virtualizing real acoustic spaces, our method should require information about the room that is as easy as possible to obtain. With this objective in mind, we show that our method produces accurate results, while only requiring the following:
\begin{enumerate}
    \item A small set of omnidirectional RIR recordings captured at sparse locations (\eg $12$), with the $xyz$ coordinates at which they were captured.
    \item The room's rough geometry, expressed as a small number of planes.
\end{enumerate}

RIRs can be easily captured by playing a sine sweep from the source location and recording it from a microphone at the listener location. In our setup, we assume a stationary audio source whose orientation and position are unknown. With this information, our goals are \cameraready{to simulate monoaural and binaural RIRs and music at arbitrary listener locations and orientations in the room.}

% \begin{enumerate}
%     \item Synthesize the omnidirectional RIR at an arbitrary listener location in the room.
%     \item Synthesize the binaural RIR at an arbitrary listener location and orientation in the room.
% \end{enumerate}
% With the synthesized dense RIR field, we can then easily simulate audio at any location of the room.

\addtocontents{toc}{\protect\setcounter{tocdepth}{1}}
\subsection{The \textbf{\name} Framework}\addtocontents{toc}{\protect\setcounter{tocdepth}{2}}

To achieve this task, we design a differentiable RIR rendering framework, dubbed \name.
\cameraready{As} an overview of the \name framework, we use the sound source and microphone location, along with the planar decomposition of the environment, to trace all specular reflection paths between the source and a listener location, up to a certain number of reflections. We estimate the sound arriving to the listener from each path using a series of parametric models for the sound source directivity and impulse response, as well as the acoustic reflection of each surface. Each model is fully differentiable, with interpretable parameters. We compute each RIR as the sum of contributions of the sound arriving from each path, combined with a learned residual. We use these models in a differentiable audio renderer to optimize parameters according to a loss function comparing our estimates to the known subset of ground-truth RIRs. We describe each model in detail below.

\subsubsection{Characterizing the Sound Source}

\myparagraph{Source Localization.}
We first estimate the location of the sound source for all subsequent steps. Based on the known subset of RIRs we use their locations and the timing of the first peak to localize the source using a traditional time-of-arrival method. More details are provided in Appendix~E.

\myparagraph{Source Directivity.}
Most real sound sources do not radiate sound uniformly in all directions. For instance, a loudspeaker will usually be much louder from the front, \cameraready{and human speakers also have distinct directivity patterns~\cite{humandirectivity}}. The source's \textit{directivity} describes the way in which the source radiates sound differently in different directions and is generally frequency dependent. For example, a loudspeaker will overall sound much louder from the front, with the higher-frequency components radiating in especially narrow beams and lower-frequency components more omnidirectionally. 
The sound source's directivity has a significant impact on the acoustic field of the room and is therefore important to model.

We model the filtering effect of exiting the sound source in any particular direction with the \textit{directivity response}. Let $\vec{d_p}$ be the absolute direction (given as a unit vector) in which the sound path exits the speaker. Our goal is to \cameraready{fit} $D(\vec{d_p})$, a function mapping $\vec{d_p}$ to a magnitude frequency response that accounts for the effect of exiting the speaker in the direction of $\vec{d_p}$. When a sound exits the speaker in the direction of $\vec{d_p}$, the frequency content of the sound wave is \textit{multiplied} by $D(\vec{d_p})$.

To model the direction-dependent frequency response, we \cameraready{fit} $F$ different heatmaps on unit spheres centered on the speaker, one heatmap for each of $F$ octave-spaced center frequencies comprising vector $\mathbf{f}$. To do this, we distribute $128$ points evenly along the surface of the unit sphere, using a Fibonacci lattice~\citep{hardin2016comparison}. We denote this set of points $L$. Let $A_{\vec{x},f_o}$ be the log-amplitude gain for sound traveling out of the speaker in the direction of $\vec{x}$ at frequency $f_o$. To determine the log-amplitude gain at $f_o$ in direction $\vec{d_p}$, we interpolate between the points on the heatmap using a spherical Gaussian weighting function, inspired by~\cite{physg2021}:
\begin{equation}
A_{\vec{d_p},f_o} =  \frac{\sum_{x \in L} A_{x,f_o} e^{-\lambda(1-\vec{d_p}\cdot x)}}{ \sum_{x \in L} e^{-\lambda(1-\vec{d_p}\cdot x)}},
\end{equation}
where $\lambda$ is a \cameraready{fixed} sharpness \cameraready{value} shared across all heatmaps. In order to obtain the full frequency response for the direction $d$, we linearly interpolate between the log-amplitude gains \cameraready{as in~\cite{hars03}}, and then exponentiate them to convert them to linear amplitude values:
\begin{equation}
D(\vec{d_p}, f_o) = \rm e^{ \ell(\mathbf{A_d}, \mathbf{f}, f_o)},
\end{equation}
where $\ell$ represents linear interpolation on the vector of decibel values $\mathbf{A_d}$ indexed by center frequencies $\mathbf{f}$, based on query frequency $f_o$.

\myparagraph{Source Impulse Response.} Since the room impulse response relates the source signal fed to the speaker to the sound heard in the room, we must also account for the way that the source modifies the source signal being fed to it. For instance, if the source is a loudspeaker, it may attenuate or boost certain frequencies. We model these effects by learning a source impulse response $IR_s$ in the time domain, \cameraready{thus approximating the source's response as a linear system~\cite{benichouxsource}} and convolving it with our RIR.

% Given a guess for the source's location, we compute the distance between the source location and each of the listener locations. Using these distances, 

% Second, we threshold each room impulse response to determine the first peak of the room impulse response. Second, 

% We show that, within the 

% \subsection{Differentiable RIR Renderer} \label{sec:diffrir}
% We center our model around a differentiable RIR renderer, which fits to a single source location $S_{xyz}$ and renders a room impulse response at any listener location $L_{xyz}$ in the room. The input to the renderer is the 3D location of the queried listener location $L_{xyz}$. The renderer learns 1) the reflection characteristics of each surface in the room, 2) the directivity pattern of the sound source, and 3) the impulse response of the sound source (the way that the sound source modifies its input signal, \eg, a loudspeaker may emphasize bass frequencies).

% We characterize each reflection path computed by our image-source tracer with 1) $S$, the set of surfaces off which it reflects, 2) $d$, a unit vector in the direction at which the path exits the speaker, and 3) $t$, the path's time of arrival, determined by the path's length. Our RIR renderer (see ~\ref{sec:diffrir}) uses these three attributes to determine the path's distinct contribution to the rendered room impulse response.

\subsubsection{Modeling and Characterizing Reflections}
We trace each specular reflection path and model the acoustic effects of each reflection along the path, with unique reflection parameters for each surface in the environment.

\myparagraph{Reflectivity.}
When a sound wave encounters a surface, a fraction of the sound wave's energy will be specularly reflected, while the remaining energy will be absorbed, transmitted, diffusely reflected, or diffracted. These effects vary by frequency, depending on the texture and material properties of each surface.

For each surface $s$, we \cameraready{fit} a vector $\mathbf{V_s}$ of $F$ different values representing the magnitude of sound specularly reflected by the surface at each of $F$ octave-spaced centered frequencies in vector $\mathbf{f}$. We apply the sigmoid function to these values to determine the \textit{energy} reflection coefficients (the proportion of specularly reflected sound energy) at each frequency. Next, we determine the \textit{amplitude} reflection coefficients (the amount that the surface attenuates the incoming sound at each frequency in terms of linear amplitude gain) by taking the square root of the energy reflection coefficients \cameraready{~\cite{landonnoise}}. Using the amplitude reflection coefficients at the $F$ center frequencies, we obtain the amplitude gains for arbitrary frequencies through linear interpolation. This gives us the \emph{reflection response} $R_s$, a magnitude frequency response representing the surface's effect on incoming audio of different frequencies. Thus, the formula for $R_s$ is:
\begin{equation}
R_s(f_r) = \ell \left( \sqrt{\sigma{(\mathbf{V_s})}}, \mathbf{f}, f_r \right).
\end{equation}
Here, $\sigma$ denotes the sigmoid function, and $\ell$ is a linear interpolation from the coefficients $\mathbf{V_s}$ based on the relation of the query frequency $f_r$ to the center frequencies $\mathbf{f}$.

\myparagraph{Reflection Paths.}
Given the estimated source location $S_{xyz}$, a listener location $L_{xyz}$, and a planar representation of the room's geometry, we use the image-source method~\citep{allen1979image} to efficiently compute all of the specular reflection paths between the source and listener in the room, up to a particular order $N$ (\eg 5). The method considers all permutations from 1 to $N$ of these surfaces with repetition and, for each permutation, determines if there is a valid reflection path that travels from the source to the listener after reflecting specularly off of each of the surfaces in order. For each valid reflection path $p$ from source to listener, we track the length of the reflection path $l_p$, the ordered list $S_p$ of reflection surfaces along the path, and the direction from which the path exits the source $\vec{d_p}$. 

Rooms often contain parallel surfaces, which lead to prominent higher-order reflections. These reflections result in ``axial modes," which are powerful room resonances with especially long reverberation times~\citep{rindel2015modal}. Thus, in addition to computing all $N^{\text{th}}$-order reflection paths for all possible orderings of surfaces, our image-source algorithm also computes all valid reflection paths for pairs of parallel walls, up to \cameraready{a much higher order, \eg 50}. This modification, which we call \emph{axial boosting}, \cameraready{improves} the model's performance \cameraready{(see Appendix D.4)} in adversarial cases like the Hallway, with a computational overhead that scales linearly rather than exponentially with reflection order. We discuss additional surface interactions, such as diffuse reflection, in \Cref{sec:residual}.

%Note that since the wavelength of sound is much larger than the wavelength of light, the relative effect of specular reflections in real environments is significantly more prominent in sound than in light. For a given surface, the effect of specular reflection overshadows the effect of diffuse reflection at sound wavelengths that are large compared to the `grain size' of the surface's texture~\citep{paspweb2010}. For the sound wavelengths relevant to human hearing, we assume that most surfaces are smooth enough to reflect sound specularly.  \mason{this paragraph could be removed}

\subsubsection{Combining Models}
We combine these reflection and sound source models to estimate the contribution of each reflection path. We then sum the contributions across all paths and add a residual to estimate the RIR for a given source and listener location.

\myparagraph{Contribution of a Single Reflection Path.}
In summary, for each individual reflection path $p$, the outgoing direction $\vec{d_p}$ from the source, the ordered list $S_p$ of reflected surfaces, and the total path length $l_p$ each have distinct effects on rendering the path's contribution. $D(\vec{d_p})$ characterizes the frequency response of the source from the path's outgoing direction. The reflection of each surface $s \in S_p$ attenuates the amplitude of the sound in a frequency-dependent fashion parameterized by $R_s$. The total reflection-based attenuation is the product of the frequency response across all $s \in S_p$. Finally, we use the path length $l_p$ to compute the time of arrival $t_p$ by dividing the path length $l_p$ by the speed of sound. We also use $l_p$ to estimate the attenuation of the amplitude due to spherical propagation, where the amplitude is inversely proportional to $l_p$, as well as air absorption, which we characterize by air absorption coefficient $\alpha$~\citep{paspweb2010}.

Thus, the function $K$ that computes the time-domain contribution of each individual path is:
\begin{equation}
\label{eqn:onepath}
K(d_p, S_p, t_p) = \frac{\alpha^{t_p}}{\cameraready{\rho}} \tau \left[\mathcal{M} 
\left(D(d_p) \odot \prod_{s \in S_p} R_s \right) , t_p\right],
\end{equation}
where $\odot$ is the element-wise product, \cameraready{$\rho$ is the length of the reflection path in meters}, and $\tau_t$ is the time-shift operator, which delays its input signal by $t_p$ \cameraready{seconds}. $\mathcal{M}$ is a minimum-phase inverse Fourier transform, which computes a time-domain filter from a magnitude frequency response, assuming minimum phase. The minimum phase assumption can be used to approximate the phase of an acoustic reflection given a desired magnitude frequency response \citep{minphase}. More details are in Appendix~E.

\myparagraph{Modeling Residual Effects.}\label{sec:residual} For the purposes of gradient-based optimization, we require a model that is fast, simple, and differentiable. Consequently, we do not explicitly model many physical phenomena, including diffuse reflection, diffraction, transmission, refraction, and higher-order specular reflections. Modeling all of these effects would increase our model's computational footprint, impeding the iterative process of fitting to a real scene. Instead, we approximate these effects as spatially uniform, with some theoretical justifications. As the reflection order increases, the number of reflection paths grows exponentially, making individual reflections less distinguishable. This comprises a sound field that, in real rooms, is approximately uniform and isotropic~\citep{Paxton2017roombook, Nolan2020diffuse, jeong2016a}. Diffuse reflections in particular can contribute to the uniformity of the sound field~\citep{visentin2020effect}. We approximate the total effect of high-order specular reflections, diffuse reflections and other effects as uniform, modeling them with a spatially-invariant residual signal $r$.

%Our ablations show the importance of fitting this residual~\ref{sec:ablations}.

\input{figures/room_photos}

\myparagraph{Overall Formula.}
Given respective source and listener locations $S_{xyz}$ and $L_{xyz}$, we render the early-stage RIR by summing the contributions from all reflection paths, then convolve the result with the source's impulse response $IR_s$.
\begin{equation}
\label{eq:full_rir}
\mbox{RIR}(S_{xyz}, L_{xyz}) =  \gamma\left[IR_s \circledast \sum_{p \in P} K(d_p, S_p, t_p)\right] + (1-\gamma)r
\end{equation}

In this formula, $\circledast$ denotes convolution, and $P$ is the set of all paths between the source and listener locations. As $r$ is intended to capture higher-order reflections, its effects are likely to become more dominant later in the impulse response, whereas the traced paths are intended to characterize the early-stage reflections. For this reason, we \cameraready{fit} 16 points on a temporal spline $\gamma$ that interpolates a relative weighting between the contributions of the late-stage residual and those of explicitly computed reflection paths.

\subsubsection{Fitting and Inference}
We estimate the parameters of each acoustic model in the environment in an iterative analysis-by-synthesis process. Inspired by \cite{clarke2022diffimpact} and \cite{engel2020ddsp}, we optimize according to a multi-scale log-spectral loss comparing rendered RIR $\hat{W}$ with the ground-truth RIR $W$ measured at the same location. \cameraready{The specific loss formulation is in Eq.~6 in Appendix~E.}

For inference, we simply compute Equation~\ref{eq:full_rir} for a point at a novel location, computing all the specular paths below the maximum \cameraready{order} between the source and the novel location, etc., and using the parameters we \cameraready{determined} from the analysis-by-synthesis process.

\myparagraph{Binauralization.}\label{sec:binauralization}
We train our model on single-channel RIRs recorded using omnidirectional microphones. However, immersive spatial audio requires binauralization - the process of converting single-channel audio into left and right channels, in a way that mimics human perception. The shape of the head, the acoustic shadow it casts, and the differences in time-of-arrival between the left and right ears all result in distinct perceptual cues that help place the listener in the scene~\cite{geronazzo2018impact, yao2017headphone}. These effects are typically modeled by head-related impulse responses (HRIRs). There is a different HRIR for each incoming audio direction. To render binaural audio, the incoming audio \cameraready{from each reflection path} is convolved with an HRIR sampled from the SADIE II dataset~\citep{armstrong2018perceptual} corresponding to its incoming direction. This allows our model to approximate perceptually accurate binaural audio, which captures the effects of the human head, with merely monaural supervision.

%% file: figures/room_photos.tex
\begin{figure*}[t]
    \centering
    \includegraphics[width=\linewidth,trim={0 0.5cm 0 0.5cm}, clip]{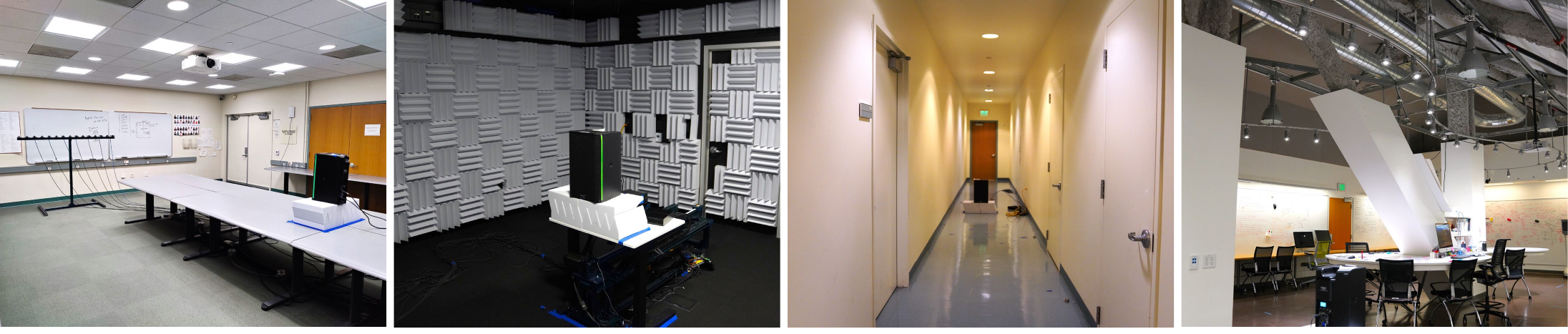}
    \vspace{2pt}
    \begin{minipage}{0.25\linewidth}
        \centering
        \caption*{(a) Classroom}
    \end{minipage}%
    \begin{minipage}{0.25\linewidth}
        \centering
        \caption*{(b) Dampened Room}
    \end{minipage}%
    \begin{minipage}{0.25\linewidth}
        \centering
        \caption*{(c) Hallway}
    \end{minipage}%
    \begin{minipage}{0.25\linewidth}
        \centering
        \caption*{(d) Complex Room}
    \end{minipage}
    \vspace{-22pt}
    \caption{Photos of each room used for the \name Dataset, each shown in its base configuration.}
    \label{fig:DatasetPictures}
    % \vspace{-10pt}
\end{figure*}

%% file: 3_dataset.tex
\addtocontents{toc}{\protect\setcounter{tocdepth}{0}}
\section{The \name Dataset}
\addtocontents{toc}{\protect\setcounter{tocdepth}{2}}
\label{sec:datasets} 
%We collect a novel dataset of real monoaural and binaural RIRs and music recordings  
%in four different rooms, as illustrated in Figure~\ref{fig:DatasetPictures}.}
To evaluate methods of rendering and interpolating RIRs, we collect \cameraready{a novel dataset} of real \cameraready{monoaural and binaural} RIRs and music data in four different rooms, as illustrated in Figure~\ref{fig:DatasetPictures}.
Table~\ref{tab:Datasets} further summarizes the dimensions and reverberation time measurements of each room.
In particular, we choose the following rooms to represent a wide range of room layouts, sizes, geometric complexities, and reverberation effects:
\begin{enumerate}
    \item \textbf{Classroom.} A standard classroom with 13 rectangular tables combined into three groups, a chalkboard, two whiteboards, drywall walls, a carpeted floor, office tile ceiling, and three doors. There is ventilation noise.
    \item \textbf{Dampened Room.} A semi-anechoic chamber with a carpeted floor, all four walls covered with jagged acoustic foam wedges, and specialty acoustic tile ceiling.
    \item \textbf{Hallway.} A narrow, highly reverberant hallway, with two wooden doors, a tile floor, and drywall ceiling and walls.
    \item \textbf{Complex Room.} A room with an irregular shape that resembles a pentagonal prism. Portions of the side wall and ceiling are covered with acoustic panels. There are three pillars in the middle of the room, one slanted diagonally. A portion of the rear wall is glass which is internally covered with paper posters. There are 7 tables, one of which is in a figure-eight shape. There are exposed air ducts, six hanging lights, water pipes, monitors, and chairs, as well as various large objects, such as a shelf. There is significant ventilation noise.
\end{enumerate}
\vspace{0.5em}
% Images of these rooms are shown in Figure~\ref{fig:DatasetPictures}, and 
% Table~\ref{tab:Datasets} provides the dimensions and reverberation time measurements of each room.

To collect audio recordings, we place a QSC K8.2 Loudspeaker in a particular location and orientation in the room and play sine sweeps to measure real RIRs in several hundred precisely-measured listener locations using a custom-built microphone array. In addition, we play and record several 10-second music clips selected from the Free Music Archive dataset~\citep{defferrard2016fma} from the same listener and speaker locations. The music and RIRs are recorded using multiple time-synchronized Dayton Audio EMM6 omnidirectional microphones, as well as a 3Dio FS XLR microphone, which features ear-shaped silicone microphones to model human hearing and captures binaural audio. 

\myparagraph{Additional Configurations.} We also collect additional subdatasets in some rooms where we slightly modify each room configuration. In each such subdataset, we vary the location and/or orientation of the speaker, or the presence and location of standalone whiteboard panels in the room. \cameraready{We use these additional configurations to evaluate zero-shot virtual speaker rotation and translation, and panel insertion and relocation. We include these evaluations and details on these configurations in Appendix~C. While previous RIR datasets include varying room configurations~\cite{wang2023soundcam,mckenzie2021arniRIR,gotz2021ambisonics} the \name Dataset is the first to our knowledge that also includes monoaural and binaural music recordings.}

\begin{table}
\small
\centering
\begin{tabular}{lcccc}
\toprule
Room & Size (m) & RT60 (s) & \# of Points\\
\midrule
Classroom & $7.1 \times 7.9 \times 2.7$ & 0.69 &  630 \\
Dampened & $4.9 \times 5.2 \times 2.7$ & 0.14 &  768 \\
Hallway & $1.5 \times 18.1 \times 2.8$ &  1.41 &  936 \\
Complex &  $8.4 \times 13.0 \times 6.1$ & 0.78 &  672 \\
\bottomrule
\end{tabular}
\caption{Characteristics of each room and corresponding subdataset. The last column is the number of distinct microphone-speaker location pairs for which both RIRs and music are recorded, across all configurations. RT60 reverberation times are each room's average across frequencies and sub-configurations. \cameraready{For the Complex room, the size of its bounding box is reported.}}
\label{tab:Datasets}
\vspace{-10pt}
\end{table}

%% file: 5_experiments.tex
\addtocontents{toc}{\protect\setcounter{tocdepth}{0}}
\section{Experiments}
\addtocontents{toc}{\protect\setcounter{tocdepth}{2}}

For each room in our collected dataset, we evaluate our performance on the tasks of \cameraready{rendering both omnidirectional RIRs and music at unseen listener locations.} In each room configuration, we select 12 omnidirectional RIRs to train our model. We then use our model to render RIRs at unseen locations in the test set, and compare our rendered RIRs to the ground-truth RIRs using metrics we detail in Section \ref{p:metrics}. \cameraready{To simulate music playing in the room, we convolve our rendered RIRs with five different source music files, and compare the result to real recordings of the same music files being played in the room, across the same metrics.}

%To evaluate our method of binauralization, we collect binaural RIRs at several locations in all rooms except for the classroom, using our 3Dio binaural microphone. We compare the RIRs we binauralize from single-channel audio to ground-truth binaural RIRs recorded from our binaural microphone, at the same location.

\myparagraph{Baselines.}
We compare our method with nearest neighbor (NN) and linear interpolation baselines, which are widely used to interpolate RIRs~\cite{luo2022learning, richard2022deep, chen2023novel}. 
We also compare with Deep Impulse Response (DeepIR)~\cite{richard2022deep} and Neural Acoustic Fields (NAF)~\cite{luo2022learning}, which are both deep-neural-network-based (DNN-based) frameworks. 
DeepIR predicts the monaural RIR at novel locations based only on the location's coordinates, while NAF uses the location combined with local geometric features to estimate the RIR. 
In addition, NAF was originally designed for binaural rendering.
Thus, we modify NAF to output monaural audio for the monaural RIR estimation task.
We also compare our method with Implicit Neural Representation for Audio Scenes (INRAS)~\cite{su2022inras}, which uses a combination of DNNs to more explicitly model specular and diffuse reflections at a subset of points in a scene's 3D mesh.% In order to generate binaural audio from baseline methods that output monoaural audio, we render monoaural audio at the location of each ear of the 3Dio binaural microphone, and then combine the two channels.

Additional details on baselines and any necessary adjustments we made to them are included in Appendix~F.
% To evaluate binaural rendering, we used the trained monaural NAF to render left and right channels one-by-one resulting in binaural rendering.
%Almost other related methods such as AV-NeRF \cite{liang2023avnerf} and NVAS \cite{chen2023novel} tend to require well pre-processed visual features; the videos need to be recorded at all audio points, and both of its videos and the corresponding recorded audio should be synchronized.

% We use real data for two reasons: First, our method is motivated by the task of virtualizing real acoustic spaces. Second, any method of room impulse response interpolation requires generating a room impulse response. Using real data instead of simulated data ensures our model can be used for virtualizing real rooms, instead of learning to replicate the generation process of the simulator.

% We collect the data using a custom-built microphone array frame designed to accommodate 12 omnidirectional measurement microphones. We move the frame to several hundred precisely-measured xyz locations. 
\addtocontents{toc}{\protect\setcounter{tocdepth}{1}}
\subsection{Results}
\addtocontents{toc}{\protect\setcounter{tocdepth}{2}}

\input{tables/compare}
\input{tables/ablations}
\myparagraph{Metrics.}\label{p:metrics}
We compare rendered audio to ground-truth audio using two metrics: 
\begin{enumerate}
    \item \textbf{Multiscale Log-Spectral L1 (\textbf{Mag})}.   
    A comparison of rendered and GT waveforms in time-frequency domain at multiple temporal and frequency resolutions~\cite{clarke2022diffimpact,engel2020ddsp}.
    \item \textbf{Envelope Distance (\textbf{ENV})}.
    The L1 distance between the log-energy envelopes of the ground-truth and rendered waveforms. Energy decay envelopes are used to extract the decay curve of the RIR, which characterizes the room's reverberant qualities~\citep{energyenv1}.  We compute the signal's energy envelope by taking the envelope of the squared signal~\citep{Boashash2015}. ~\citet{satoh1998reverberation} directly use this log-energy (squared) envelope of an RIR to measure the room's RT60 reverberation time, which is a common way of characterizing the room's acoustics~\citep{RT60Energy}.
\end{enumerate}

\myparagraph{Analysis.} Our results for the base monaural prediction task are shown in Table~\ref{tab:base_results}. For the monaural prediction task, our model significantly outperforms all baselines on our metrics, across all rooms. Results for the binaural prediction task are shown in Appendix~D.1.% Note that it is difficult to compare a binaural RIR recorded from our binaural microphone with a binauralized audio originally recorded from a different microphone. It is possible our model renders high-quality, binauralized audio that performs worse on these metric because it fails to match the characteristics of our binaural microphone. Because of this, we include qualitative binauralization results in the supplementary video.

\addtocontents{toc}{\protect\setcounter{tocdepth}{1}}
\subsection{Interpretability}
\addtocontents{toc}{\protect\setcounter{tocdepth}{2}}

\label{sec:Interpretability}
We show the physically interpretable parameters our model learns for the source's directivity and reflection coefficients.
\input{figures/interpretability}

\myparagraph{Directivity Maps.} \cameraready{The left side of Figure~\ref{fig:interpretability} shows the source directivity heatmaps at various frequencies, learned from 12 training points in the Classroom subdataset.} The area near the front of the speaker emits the loudest sound across most frequencies, as expected. The figures also confirm that higher frequencies are more directionally emitted than lower ones, evident in the narrowing yellow directivity ``beam" with increasing frequency. Additionally, the fact that higher frequencies are typically emitted by the loudspeaker's tweeter at the top front of the speaker, is reflected in our heatmaps, where the yellow regions appear above the speaker's center for higher frequencies.

\myparagraph{Reflection Amplitude Responses.} The right side of Figure~\ref{fig:interpretability} shows the specular reflection amplitude responses that our model fits to some surfaces in the \cameraready{Classroom and Dampened Room}. Our model correctly infers that the carpeted floor seems to be more absorptive than the wall, which consists of more rigid and smooth materials. \cameraready{The wall in the Dampened Room is even more absorptive, as our model predicts nearly no reflection above 2 kHz.}

\input{figures/virtual_modifications}

\myparagraph{Virtual Rotation and Translation.} 
Since our model learns physically interpretable parameters, we can simulate changes to the room layout that are unseen in the training data. In \Cref{fig:virtual_modifications}, we train our model on the Dampened subdataset, and use it to simulate virtual speaker rotation and translation. We visualize these changes by plotting RIR loudness heatmaps. Since the \name Dataset also includes real data where the speaker is rotated or translated, we include quantitative evaluations on virtual speaker rotation and translation in the Appendix~C.3, as well as evaluations on virtual panel insertion and relocation.

% In addition, the smooth and flat wall appears to reflect higher frequencies relatively more than the coarsely-textured tiles on the ceiling. This also matches expectations, as rougher surfaces tend to diffuse sound instead of reflecting it, especially at higher frequencies where the `grain size' of the texture is comparable to the wavelength of the sound~\citep{paspweb2010}.

\addtocontents{toc}{\protect\setcounter{tocdepth}{1}}
\subsection{Ablation Study} \label{sec:ablations}
\addtocontents{toc}{\protect\setcounter{tocdepth}{2}}

We ablate three major components of our model (the residual, modeling the source's directivity, and modeling the source's impulse response) to determine their individual contributions.
Table~\ref{tab:ablations} shows our results. The results suggest that these components are all necessary for effectively rendering accurate RIRs at novel locations. More ablations experiments are in Appendix D.4.

\cameraready{
\addtocontents{toc}{\protect\setcounter{tocdepth}{1}}
\subsection{Additional Experiments and Visualizations.}
\addtocontents{toc}{\protect\setcounter{tocdepth}{2}}
Along with additional RIR loudness maps, Appendix B.2 shows that our model can reconstruct the modal structure of the soundfield at a low frequency. In Appendix D.2, we show that our model trained on 6 points outperforms all baselines trained on 100 points. Appendix D shows that our model is robust to geometric distortions and experiments with modeling the effects of transmission.
}

%% file: tables/compare.tex
\begin{table*}
\centering
\resizebox{17.5cm}{!}{
\begin{tabular}{ l cccc cccc cccc cccc }

    \toprule
    
    & \multicolumn{4}{c}{\textbf{Classroom}} & \multicolumn{4}{c}{\textbf{Dampened Room}} & \multicolumn{4}{c}{\textbf{Hallway}} & \multicolumn{4}{c}{\textbf{Complex Room}} \\
    
    \cmidrule(lr){2-5} \cmidrule(lr){6-9} \cmidrule(lr){10-13} \cmidrule(lr){14-17}

    & \multicolumn{2}{c}{RIR} & \multicolumn{2}{c}{Music} & \multicolumn{2}{c}{RIR} & \multicolumn{2}{c}{Music} & \multicolumn{2}{c}{RIR} & \multicolumn{2}{c}{Music} & \multicolumn{2}{c}{RIR} & \multicolumn{2}{c}{Music} \\
    
    \cmidrule(lr){2-3} \cmidrule(lr){4-5} \cmidrule(lr){6-7} \cmidrule(lr){8-9} \cmidrule(lr){10-11} \cmidrule(lr){12-13} \cmidrule(lr){14-15} \cmidrule(lr){16-17}
    
    & Mag & ENV & Mag & ENV & Mag & ENV & Mag & ENV & Mag & ENV & Mag & ENV & Mag & ENV & Mag & ENV \\ \midrule 
    
    NN & 5.99 & 1.10 & 2.95 & 1.42 & 1.36 & 0.61 & 1.99 & 1.36 & 10.14 & 3.04 & 2.62 & 1.32 & 5.52 & 0.99 & 2.39 & 1.42\\
    %OLD Linear & 7.70 & 2.13 & 4.12 & 2.42 & 1.79 & 0.77 & 2.95 & 2.29 & 13.97 & 5.96 & 3.65 & 2.19 & 7.20 & 2.00 & 3.31 & 2.20\\
    Linear & 6.44 & 1.52 & 3.34 & 1.82 & 1.55 & 0.652 & 2.43 & 1.66 & 11.63 & 4.49 & 3.11 & 1.75 & 6.03 & 1.43 & 2.74 & 1.74\\
    DeepIR & 9.23 & 2.81 & 3.15 & 1.65 & 3.09 & 3.41 & 3.39 & 2.22 & 15.71 & 10.34 & 2.97 & 1.47 & 8.08 & 2.80 & 2.62 & 1.65\\
    NAF & 6.36 & 1.38 & 3.32 & 1.75 & 2.00 & 0.73 & 3.38 & 1.54 & 12.26 & 3.82 & 3.13 & 1.46 & 6.10 & 1.31 & 2.87 & 1.71\\
    INRAS & 9.99 & 4.52 & 4.45 & 1.75 & 4.20 & 2.48 & 6.22 & 5.35 & 14.52 & 9.19 & 3.70 & 1.58 & 9.02 & 2.58 & 3.61 & 1.66\\
    \name (ours) & \textbf{5.22} & \textbf{0.94} & \textbf{2.71} & \textbf{1.36} & \textbf{1.21} & \textbf{0.56} & \textbf{1.59} & \textbf{1.19} & \textbf{9.13} & \textbf{2.95} & \textbf{2.59} & \textbf{1.25} & \textbf{4.86} & \textbf{0.92} & \textbf{2.25} & \textbf{1.41} \\

    \bottomrule

\end{tabular}
}

\caption{
    Experimental results on the task of predicting monaural RIRs and music at an unseen point. Lower is better for all metrics. Errors for RIRs are multiplied by 10.
}
\label{tab:base_results}
\end{table*}

%% file: tables/ablations.tex
\begin{table*}
\centering
\resizebox{17.5cm}{!}{
\begin{tabular}{ l cccc cccc cccc cccc }
    \toprule
    & \multicolumn{4}{c}{\textbf{Classroom}} & \multicolumn{4}{c}{\textbf{Dampened Room}} & \multicolumn{4}{c}{\textbf{Hallway}} & \multicolumn{4}{c}{\textbf{Complex Room}} \\ \cmidrule(lr){2-5} \cmidrule(lr){6-9} \cmidrule(lr){10-13} \cmidrule(lr){14-17}
& \multicolumn{2}{c}{RIR} & \multicolumn{2}{c}{Music} & \multicolumn{2}{c}{RIR} & \multicolumn{2}{c}{Music} & \multicolumn{2}{c}{RIR} & \multicolumn{2}{c}{Music} & \multicolumn{2}{c}{RIR} & \multicolumn{2}{c}{Music}  \\

\cmidrule(lr){2-3} \cmidrule(lr){4-5} \cmidrule(lr){6-7} \cmidrule(lr){8-9} \cmidrule(lr){10-11} \cmidrule(lr){12-13} \cmidrule(lr){14-15} \cmidrule(lr){16-17}

& Mag & ENV & Mag & ENV & Mag & ENV & Mag & ENV & Mag & ENV & Mag & ENV & Mag & ENV & Mag & ENV \\ \midrule 

\name & \textbf{5.22} & \textbf{0.94} & \textbf{2.71} & \textbf{1.36} & \textbf{1.21} & \textbf{0.56} & \textbf{1.59} & \textbf{1.19} & \textbf{9.13} & \textbf{2.95} & \textbf{2.59} & \textbf{1.25} & \textbf{4.86} & \textbf{0.92} & \textbf{2.25} & \textbf{1.41} \\ 

\:\:\:w/o Directivity Pattern & 5.47 & 0.97 & 3.02 & 1.49 & 1.64 & 0.63 & 3.02 & 1.54 & 9.98 & 3.09 & 2.98 & 1.34 & 5.13 & 0.94 & 2.45 & 1.46\\
\:\:\:w/o Source IR & 5.39 & 0.99 & 2.79 & 1.48 & 1.36 & 0.63 & 1.73 & 1.45 & 9.38 & 3.04 & 2.76 & 1.38 & 5.07 & 0.96 & 2.38 & 1.49\\
\:\:\:w/o Residual Component & 6.90 & 1.37 & 3.07 & 1.40 & 1.37 & 0.61 & 1.77 & 1.38 & 15.49 & 4.80 & 2.81 & 1.27 & 6.24 & 1.30 & 2.46 & 1.47\\
\bottomrule

\end{tabular}
}

% \vspace{-0.5em}
\caption{
    Ablation results. In each row, the ablated parameter is frozen to its initial value during training, \ie the Source IR is assumed to be an ideal impulse, the Directivity Pattern is assumed to be uniform at all frequencies, and the Residual Component is assumed to be zero.
}
\label{tab:ablations}
\vspace{-5pt}
\end{table*}

%% file: figures/interpretability.tex
\begin{figure}[h]
    \centering
    \begin{minipage}{0.37\linewidth}
        \captionsetup[subfigure]{labelformat=empty}
        \begin{subfigure}[b]{0.45\textwidth}
            \includegraphics[width=\textwidth, trim=7.69cm 7.95cm 8.73cm 8.47cm, clip]{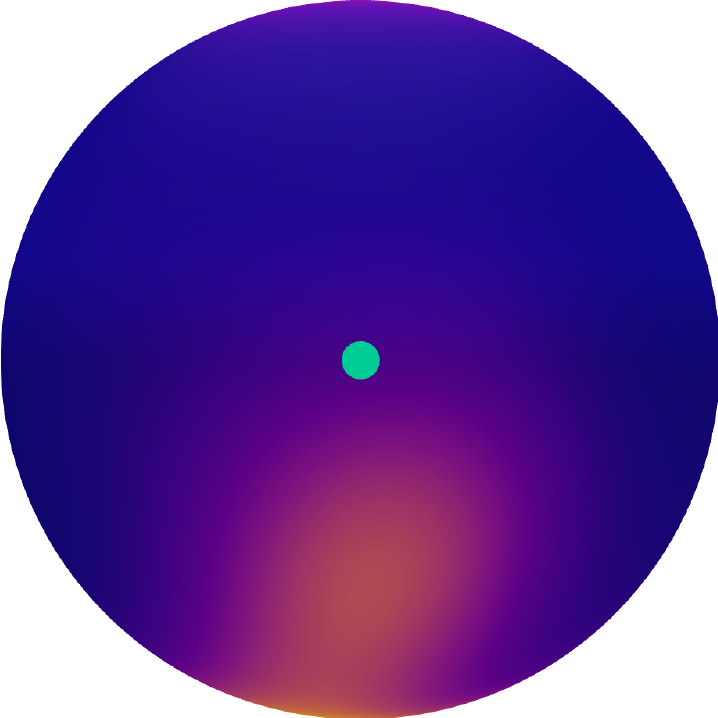}
            \caption{63 Hz}
            \label{fig:1}
        \end{subfigure}
        \hfill 
        \begin{subfigure}[b]{0.45\textwidth}
            \includegraphics[width=\textwidth, trim=7.69cm 7.95cm 8.73cm 8.47cm, clip]{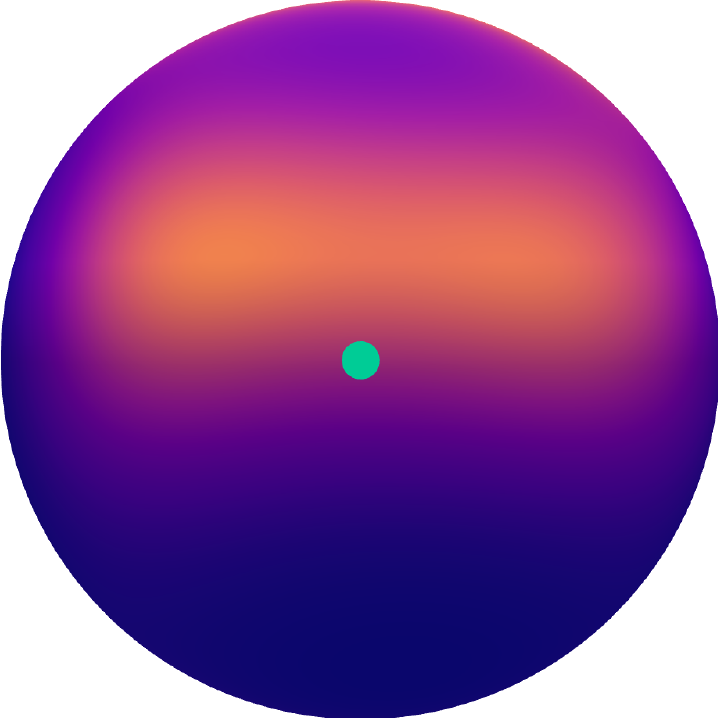}
            \caption{2000 Hz}
            \label{fig:2}
        \end{subfigure}
        \begin{subfigure}[b]{0.45\textwidth} 
            \includegraphics[width=\textwidth, trim=7.69cm 7.95cm 8.73cm 8.47cm, clip]{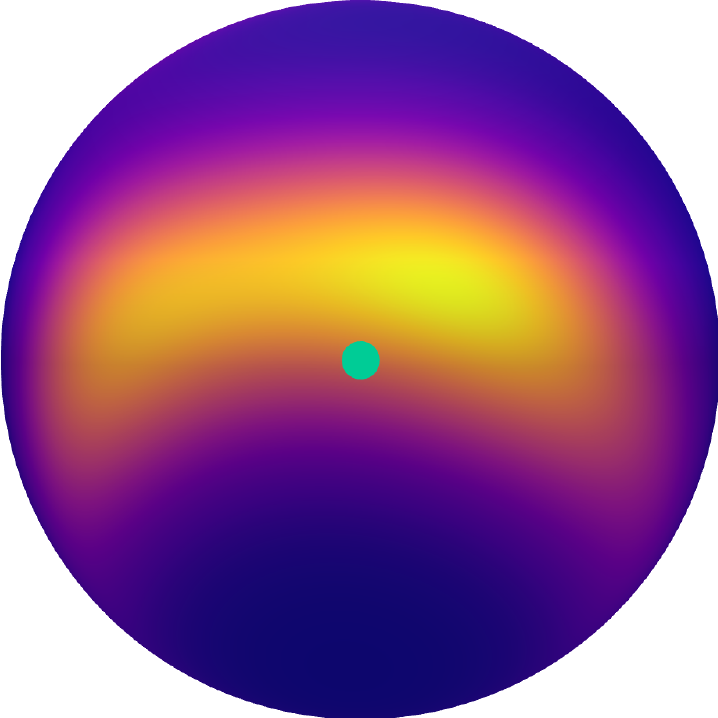}
            \caption{8000 Hz}
            \label{fig:3}
        \end{subfigure}
        \hfill
        \begin{subfigure}[b]{0.45\textwidth}
            \includegraphics[width=\textwidth, trim=7.69cm 7.95cm 8.73cm 8.47cm, clip]{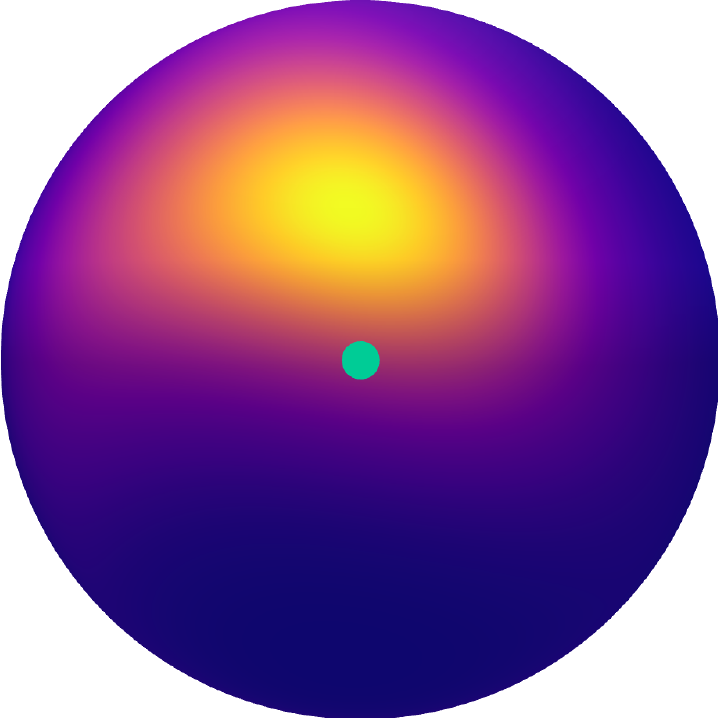}
            \caption{16000 Hz}
            \label{fig:4}
        \end{subfigure}
    \end{minipage}%
    \hfill
    \begin{minipage}{0.62\linewidth}
        \centering
        \includegraphics[width=\textwidth, trim=0cm 0.2cm 0cm 0.1cm, clip]{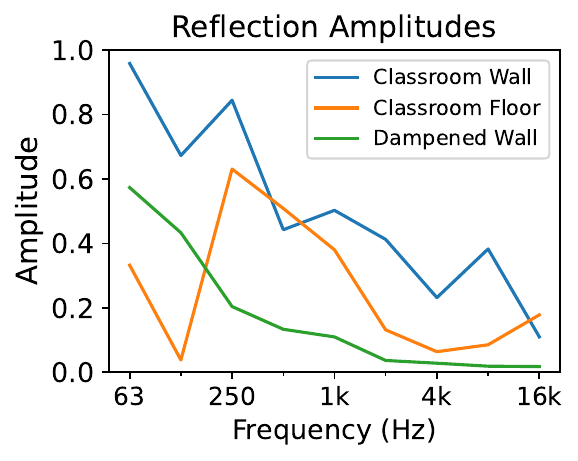}
        \label{fig:5}
    \end{minipage}
    \vspace{-1em}
    \caption{Visualization of our model's learned parameters. The left images show sample spherical heatmaps that our model fits to the speaker's directivity pattern when trained on 12 points from the Classroom subdataset. The green dot indicates the direction the speaker is facing, and the yellow regions indicate higher volume. The right image shows reflection amplitude responses that our model learns for various surfaces.}
    \label{fig:interpretability}
\end{figure}

%% file: figures/virtual_modifications.tex
\begin{figure}[h]
\centering
\vspace{-10pt}
\begin{subfigure}[b]{0.32\columnwidth}
    \centering
    \includegraphics[width=\linewidth]{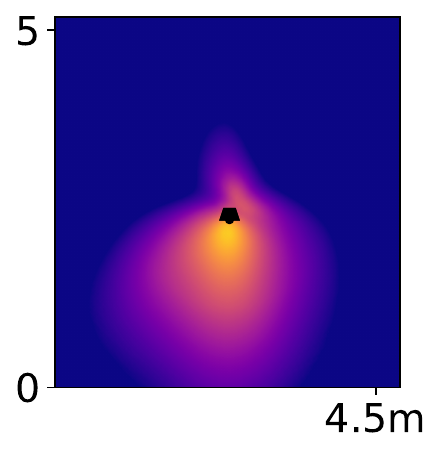}
    \caption{Base Model}
    \label{fig:image1}
\end{subfigure}
\hfill
\begin{subfigure}[b]{0.32\columnwidth}
    \centering
    \includegraphics[width=\linewidth]{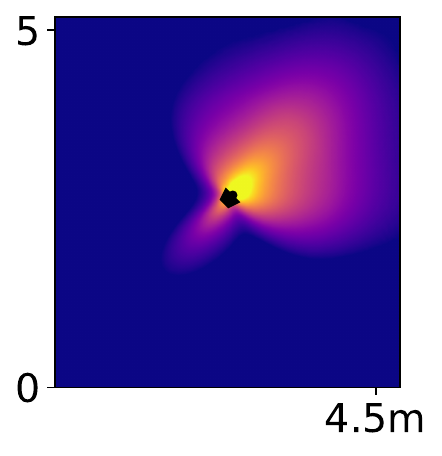}
    \caption{Virtual Rotation}
    \label{fig:image2}
\end{subfigure}
\hfill
\begin{subfigure}[b]{0.32\columnwidth}
    \centering
    \includegraphics[width=\linewidth]{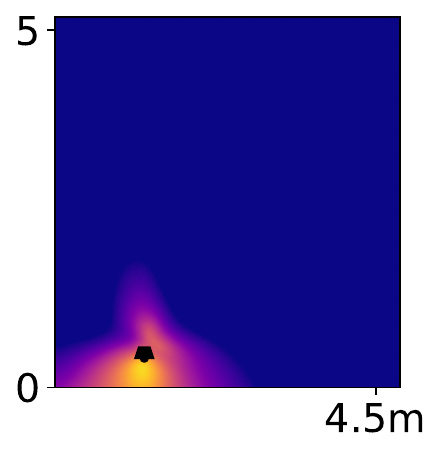}
    \caption{Virtual Translation}
    \label{fig:image3}
\end{subfigure}
\vspace{-5pt}
\caption{RIR loudness heatmaps generated from \name trained on 12 points in the Dampened Room's base subdataset.}
\label{fig:virtual_modifications}
\vspace{-5pt}
\end{figure}

%% file: 6_conclusions.tex
\addtocontents{toc}{\protect\setcounter{tocdepth}{0}}
\section{Conclusions}
\addtocontents{toc}{\protect\setcounter{tocdepth}{2}}
We presented \name, a differentiable RIR renderer capable of accurately rendering the room's acoustic impulse response at new locations, given a small set of microphone recordings and the room geometry. Future work could focus on modeling a room's acoustics implicitly by recording natural audio, thus obviating the need to measure RIRs.% In addition, our framework assumes that the room has a single sound source, which can be extended to multiple sound sources in future work.
% \ryosuke{Should we mention about dataset if it will be opened?}\sz{yes}

%% file: 7_appendix.tex
\renewcommand{\thesection}{\Alph{section}}
\setcounter{section}{0}
\setcounter{equation}{5}

\begin{figure*}
    \centering
    \includegraphics[width=\textwidth]{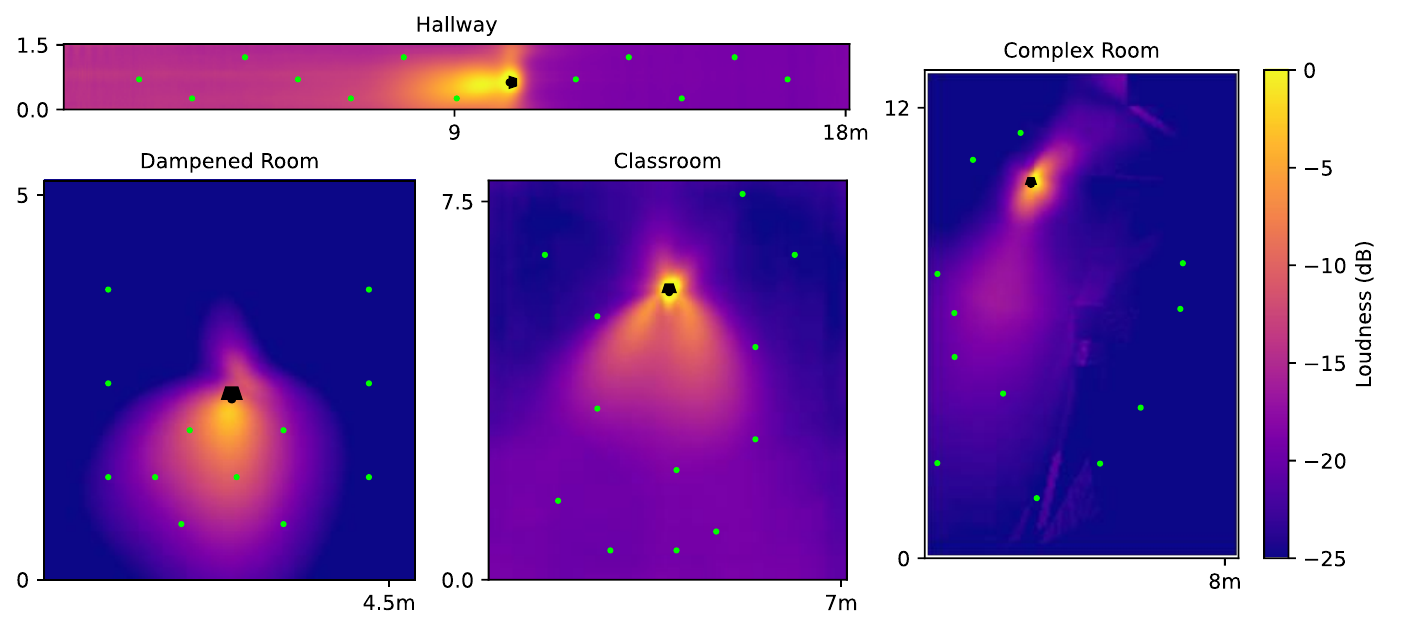}
    \vspace{-20pt}
    \caption{Visualization of RIR loudness maps generated from our model trained in each of the four base subdatasets. We measure loudness by rendering an RIR at a given listener location and measuring its RMS volume level. For each RIR rendered, we fix the height of the listener location to be 1 meter above the floor. The resolution of each xy-grid is approximately 5 centimeters in both the x and y directions. We fix the location and orientation of the speaker (indicated by the black icon) to where it was during RIR measurement. The color scale is in decibels and is consistent between rooms. The green dots indicate the xy locations of the 12 training points, which are projected onto the $z=1$ plane.
     }
    \label{fig:volumemap}
\end{figure*}

\section{Qualitative Results and Video}
Please see the supplementary video on the \href{https://masonlwang.com/hearinganythinganywhere}{website} for an in-depth qualitative analysis and comparative evaluation against baseline models.
This video showcases a simulation of a song played in two distinct environments: the Dampened Room and the Hallway. The purpose is to demonstrate the immersive quality and perceptual accuracy of the audio rendered by our model, reflecting the true characteristics of the real scenes. To achieve this, we rendered 100 room impulse responses at various locations, convolved them with the chosen source audio, and smoothly interpolated between these convolved signals. For an optimal experience of these qualitative results, we recommend using earbuds or headphones while viewing the video.

Furthermore, the video features a side-by-side comparison of our binaural audio results with those from baseline models, highlighting the enhanced realism and compelling nature of the audio generated by our model. This comparison underscores the significant qualitative improvements our model offers in creating an immersive auditory experience.
In addition, the video provides visualizations explaining our method, and the task setup.

\section{RIR Heatmap Visualizations}

\subsection{Broadband RIR Heatmaps}
After our model is trained, we can use it to visualize how the loudness of the rendered acoustic field varies spatially. To do this, we use the model trained on each of the base subdatasets to render RIRs on a dense 2D-grid of listener locations. We visualize of the root mean square (RMS) volume level of the RIRs in \Cref{fig:volumemap}, on a decibel (logarithmic) color scale. The visualizations shown are similar to those in \cite{luo2022learning, su2022inras}.

We observe several differences in the heatmaps for the different rooms. In the Dampened Room, the surfaces are less reflective, and thus, much of the soundfield's loudness is concentrated in the region in front of the speaker. This effect is reduced in the Classroom, where the soundfield is more spread out. In the Hallway, which is the most reflective room, the soundfield's volume is even more spread out, and the region behind the speaker is significantly louder than it is in any of the other rooms.

\subsection{Soundfield Reconstruction}
\begin{figure*}
\centering
\begin{subfigure}{.49\textwidth}
  \centering
  \includegraphics[width=\linewidth]{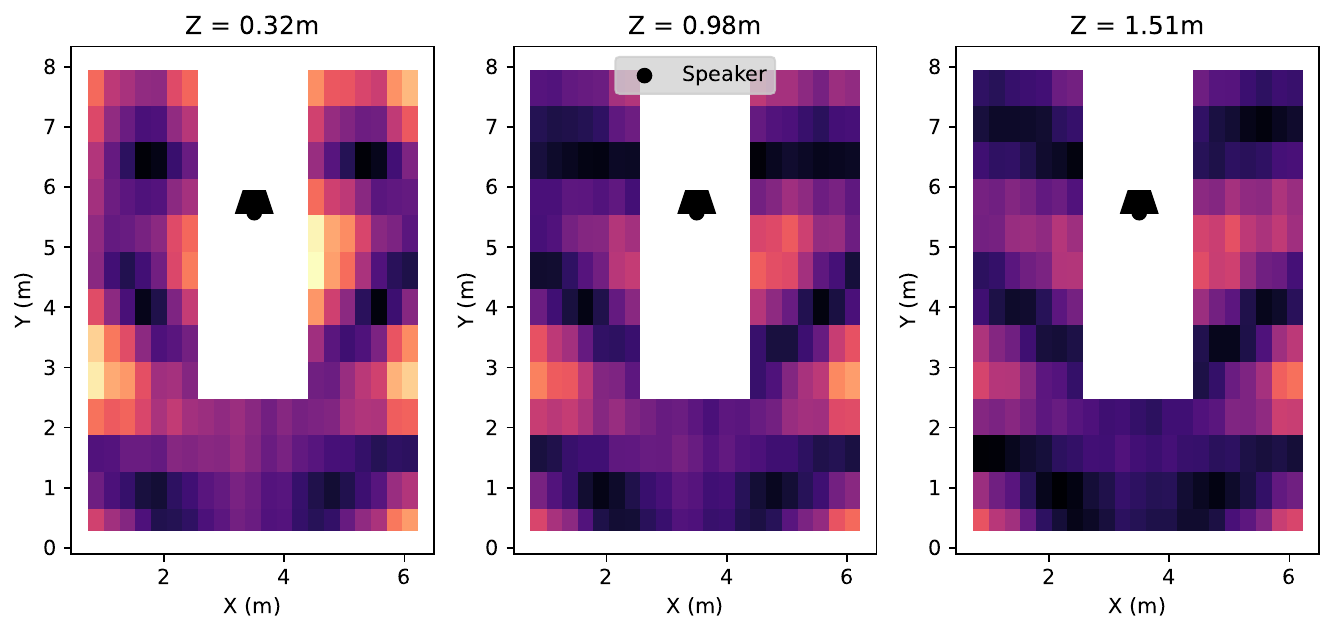}
  \caption{GT}
  \label{fig:gt}
\end{subfigure}%
\begin{subfigure}{.49\textwidth}
  \centering
  \includegraphics[width=\linewidth]{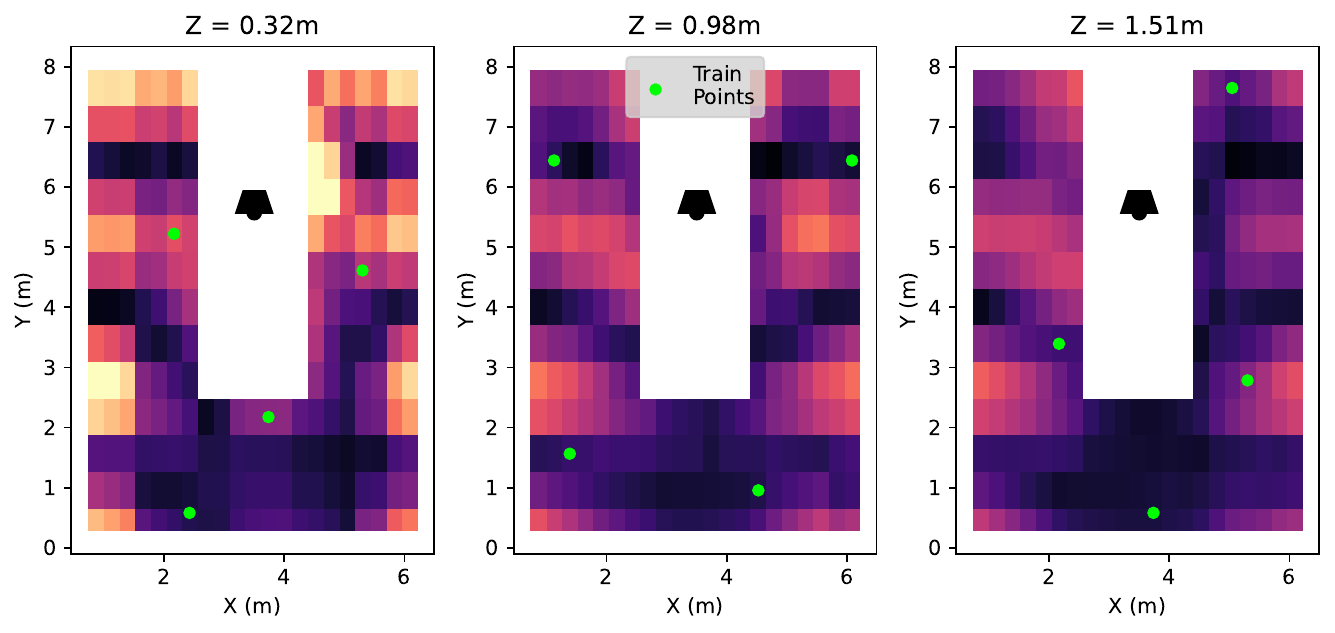}
  \caption{Ours}
  \label{fig:ours}
\end{subfigure}%
\vspace{12pt}

\begin{subfigure}{.49\textwidth}
  \centering
  \includegraphics[width=\linewidth]{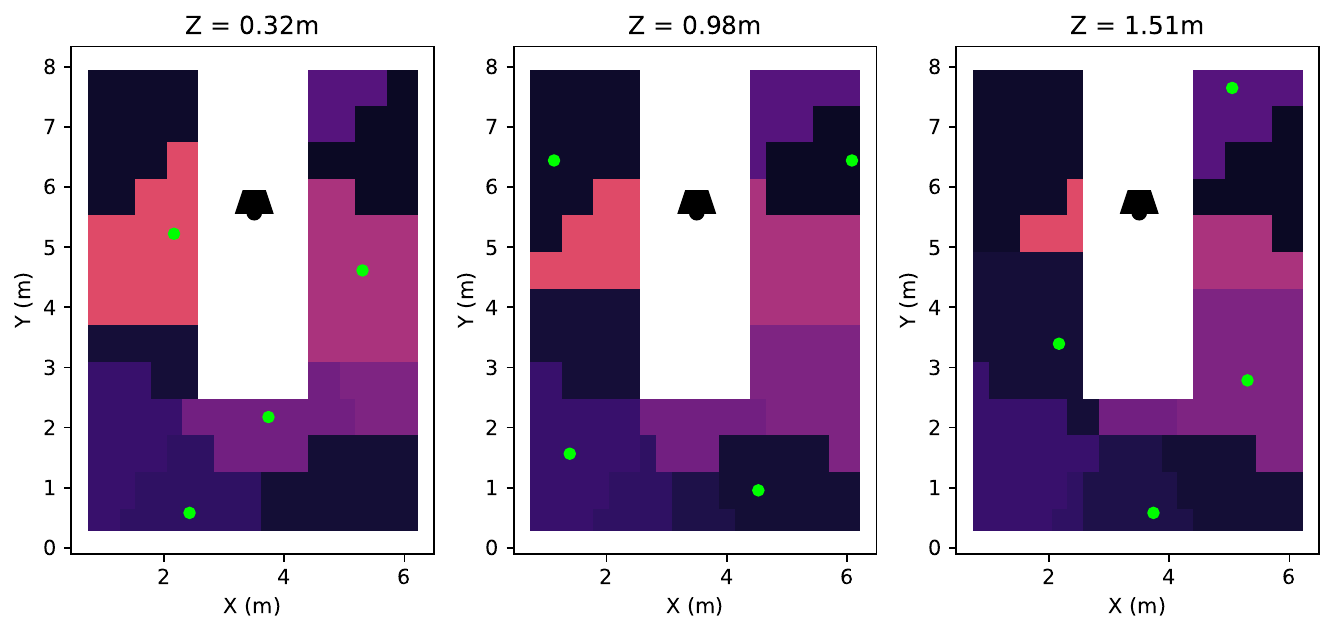}
  \caption{Nearest Neighbors}
  \label{fig:nn}
\end{subfigure}
\begin{subfigure}{.49\textwidth}
  \centering
  \includegraphics[width=\linewidth]{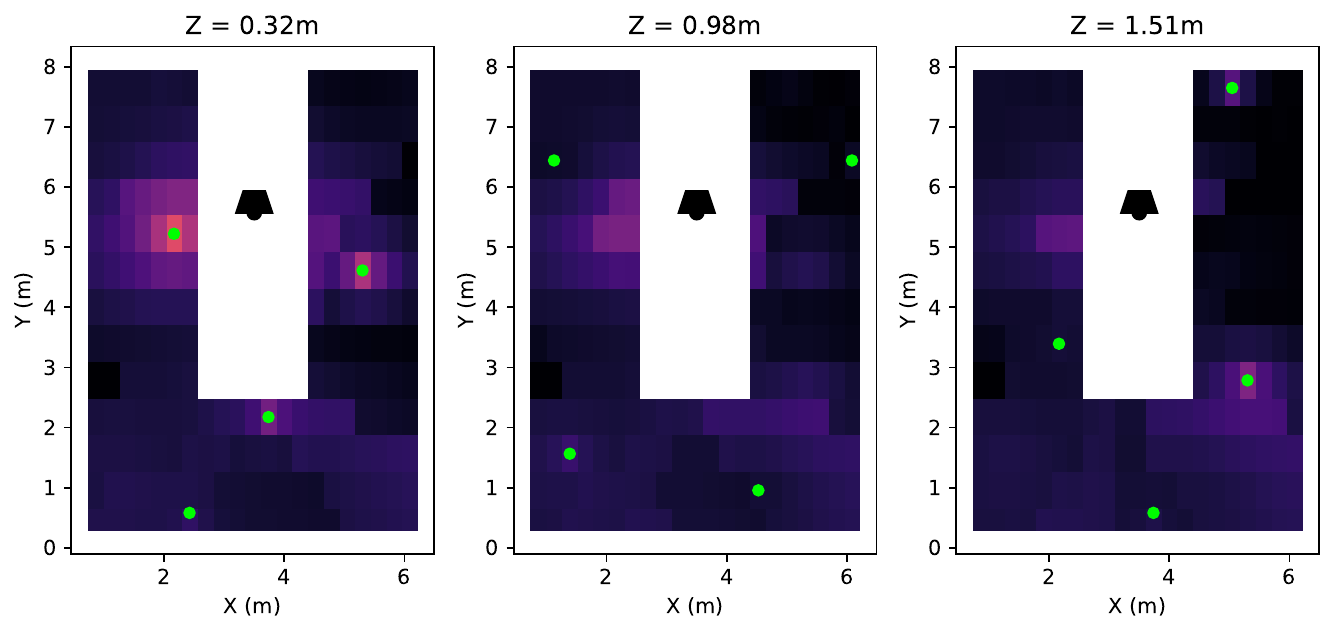}
  \caption{Linear}
  \label{fig:linear}
\end{subfigure}%
\vspace{12pt}

\begin{subfigure}{.49\textwidth}
  \centering
  \includegraphics[width=\linewidth]{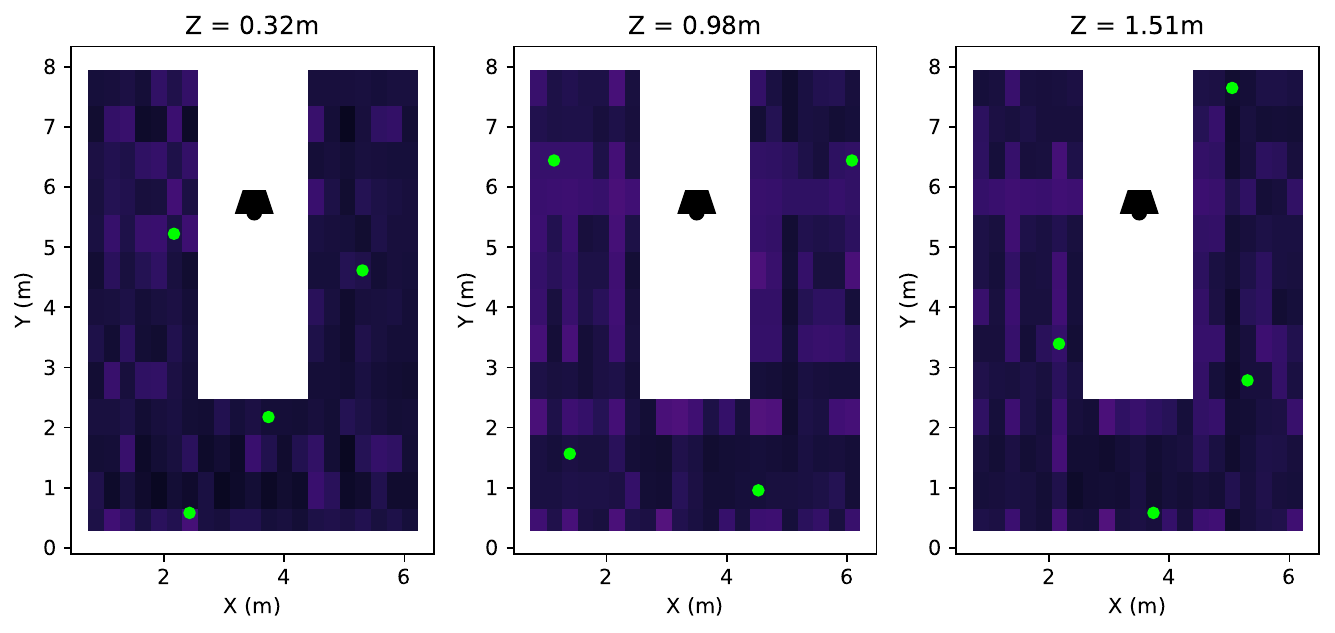}
  \caption{NAF}
  \label{fig:naf}
\end{subfigure}%
\begin{subfigure}{.49\textwidth}
  \centering
  \includegraphics[width=\linewidth]{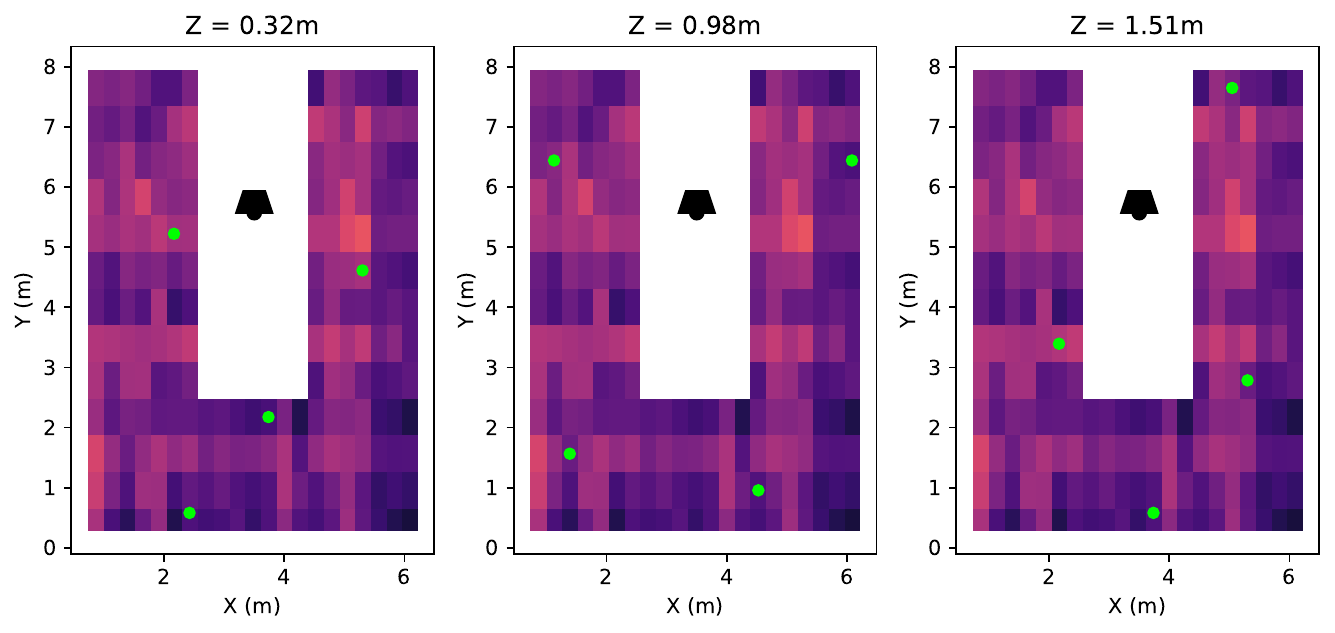}
  \caption{INRAS}
  \label{fig:inras}
\end{subfigure}
\vspace{12pt}

\begin{subfigure}{.49\textwidth}
  \centering
  \includegraphics[width=\linewidth]{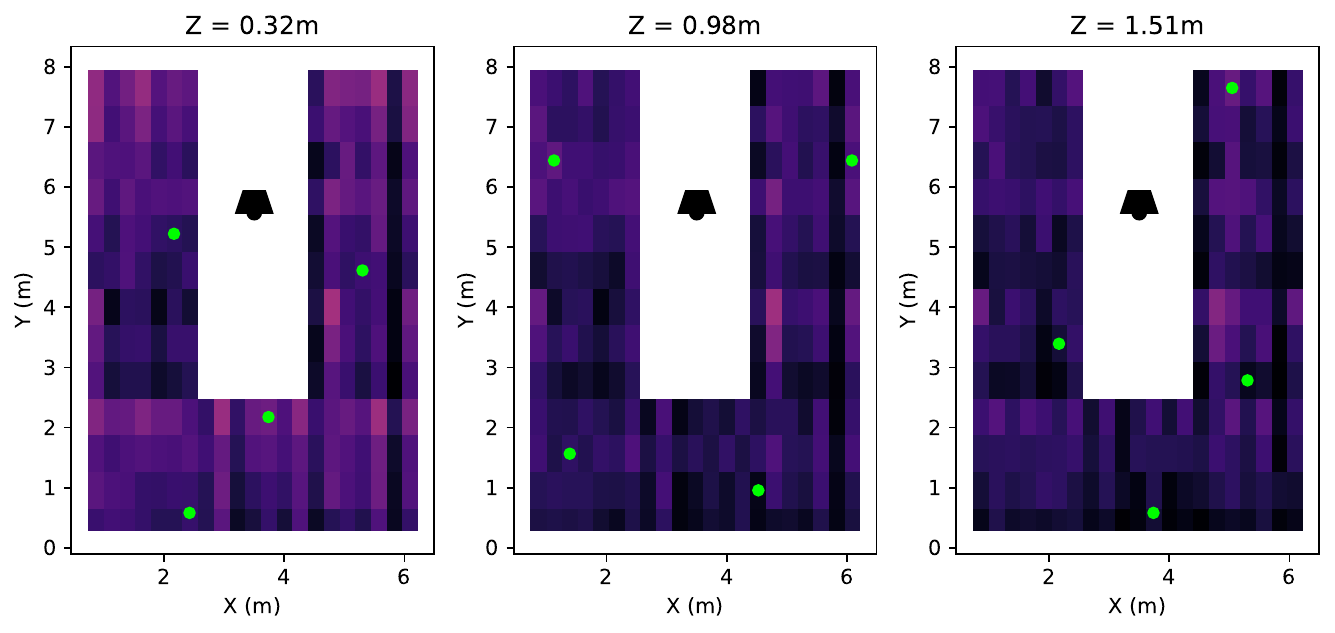}
  \caption{DeepIR}
  \label{fig:deepir}
\end{subfigure}
\caption{Visualization of RIR loudness at 70 Hz in the Classroom subdataset. The sound field intensity at a given location is measured by filtering the ground-truth or predicted RIR around 70 Hz using a 2nd order Butterworth filter~\cite{Butterworth1930} and measuring the RMS volume level of the filtered signal. Subfigure a) shows the intensity of the 70hz sound field at all locations in the subdataset. Subfigure b) shows predicted intensities at these same locations using our model trained on 12 points. We indicate the spatial locations of these 12 training points with green dots, and the speaker's location and orientation with a black icon. Subfigures c) through g) show the sound field intensity as predicted by each of our baseline models. Note that in subfigure d), the Linear baseline underestimates the soundfield intensity at locations far away from the training locations, since the linear interpolation at these locations is a weighted average of roughly uncorrelated signals whose mean is roughly zero.
}       
\label{fig:70hzreconstruction}
\end{figure*}

When observed at a single frequency, the spatial variations in sound pressure for a given sound field often exhibit modal patterns. Reconstructing the pressure levels of a sound field from a sparse set of observations is a problem of longstanding theoretical and practical interest \cite{soundfieldreconstruction, Ajdler06plenacoustic,CCRMAsoundfieldreconstruction, gaussianrecon}. Using the RIRs measured in the Classroom subdataset, we calculate the sound pressure level at 70 Hz at all locations in our subdataset, plotted in \Cref{fig:gt}. We also use the predicted RIRs from each method to predict the sound pressure level at 70 Hz at every spatial location. We find that our model learns to predict the modal structure of the RIR sound field without explicitly modeling it, while other baselines fail to do this. Note that our model approximately predicts the locations of the sound field's nodes and anti-nodes (regions of high and low intensity), even without observing training data in those locations.

\section{Results on Additional Room Configurations}\label{sec:appc}
\subsection{Description of Additional Subdatasets}
\begin{figure*}
    \centering
    \begin{tabular}{ccccc}
        \includegraphics[width=0.18\textwidth]{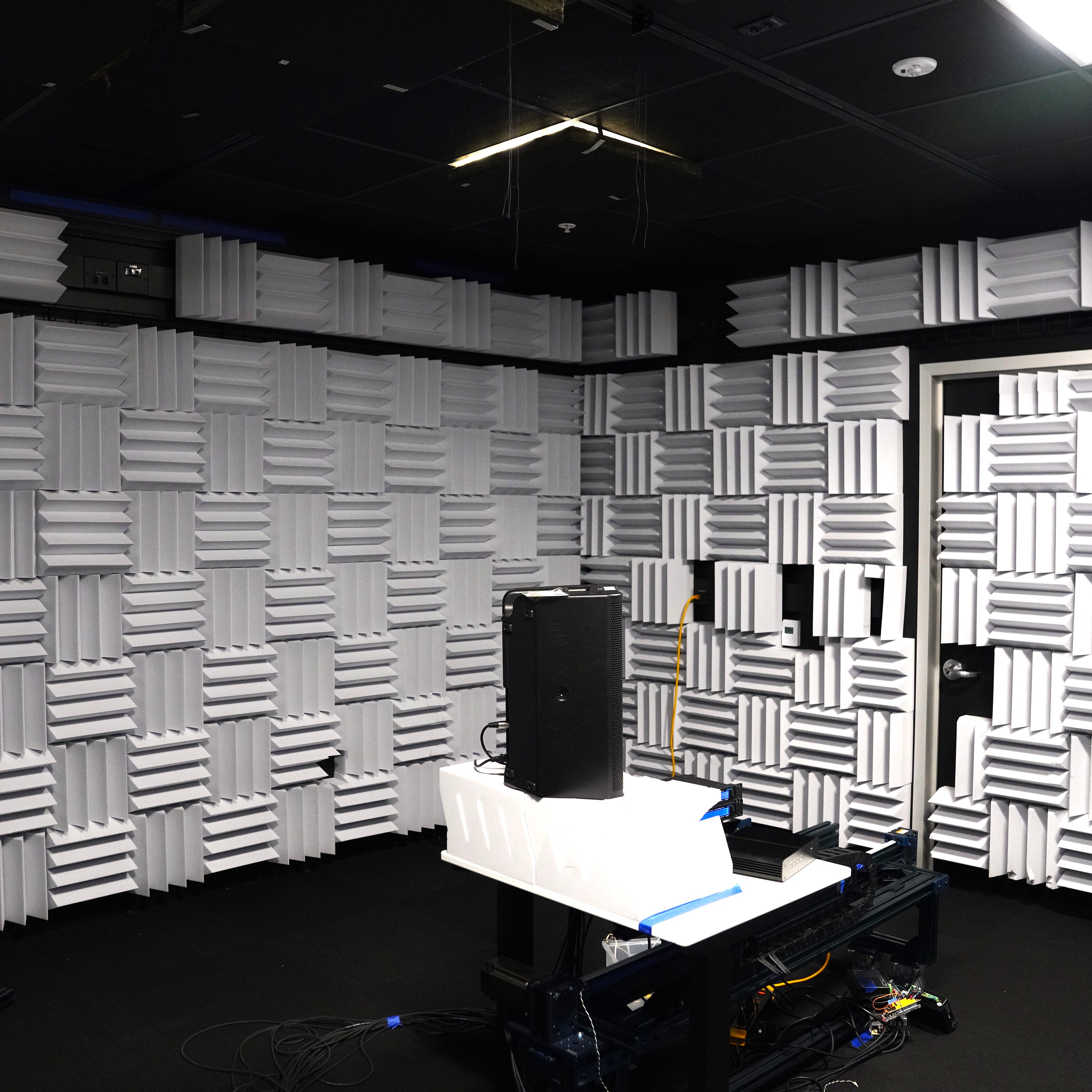} & 
        \includegraphics[width=0.18\textwidth]{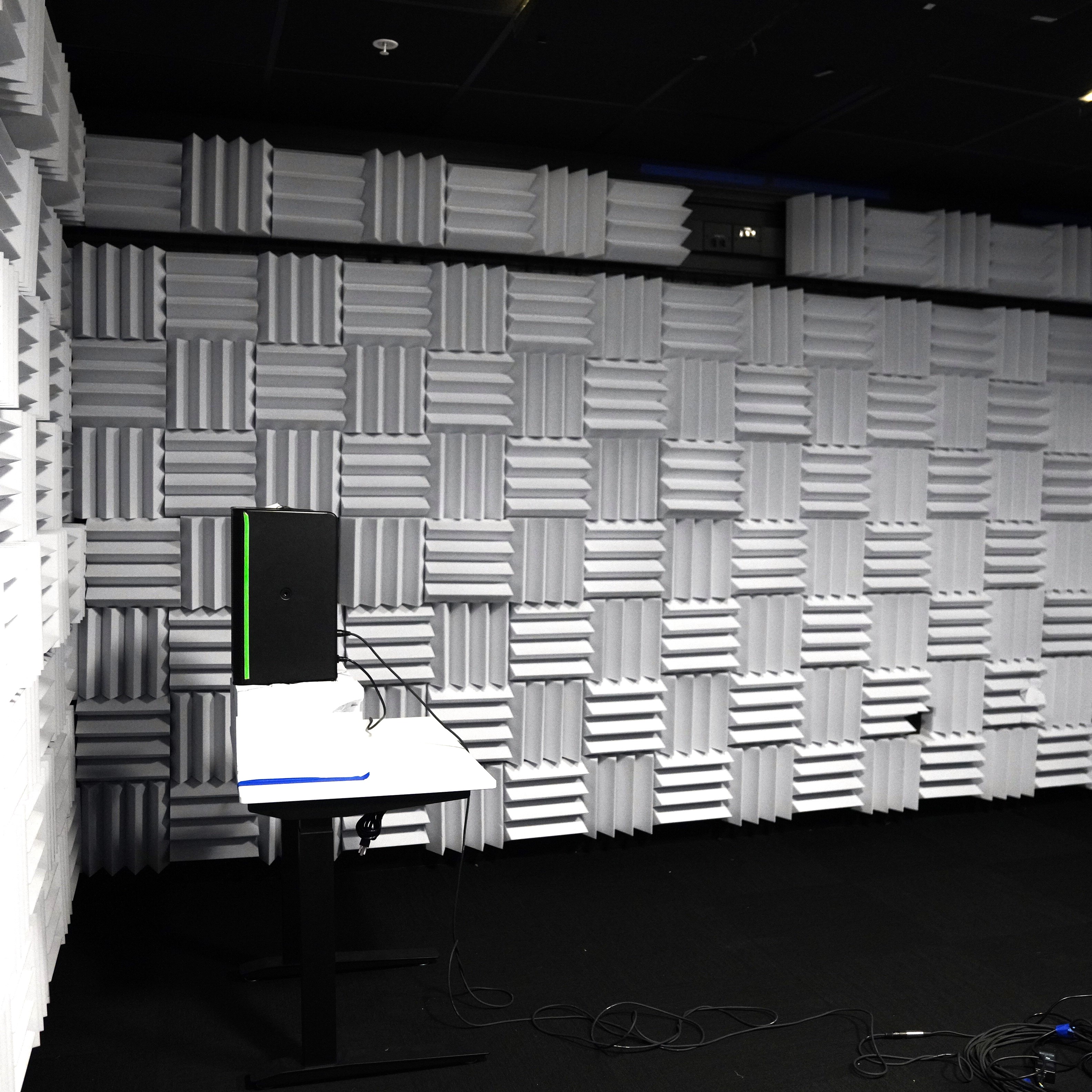} & 
        \includegraphics[width=0.18\textwidth]{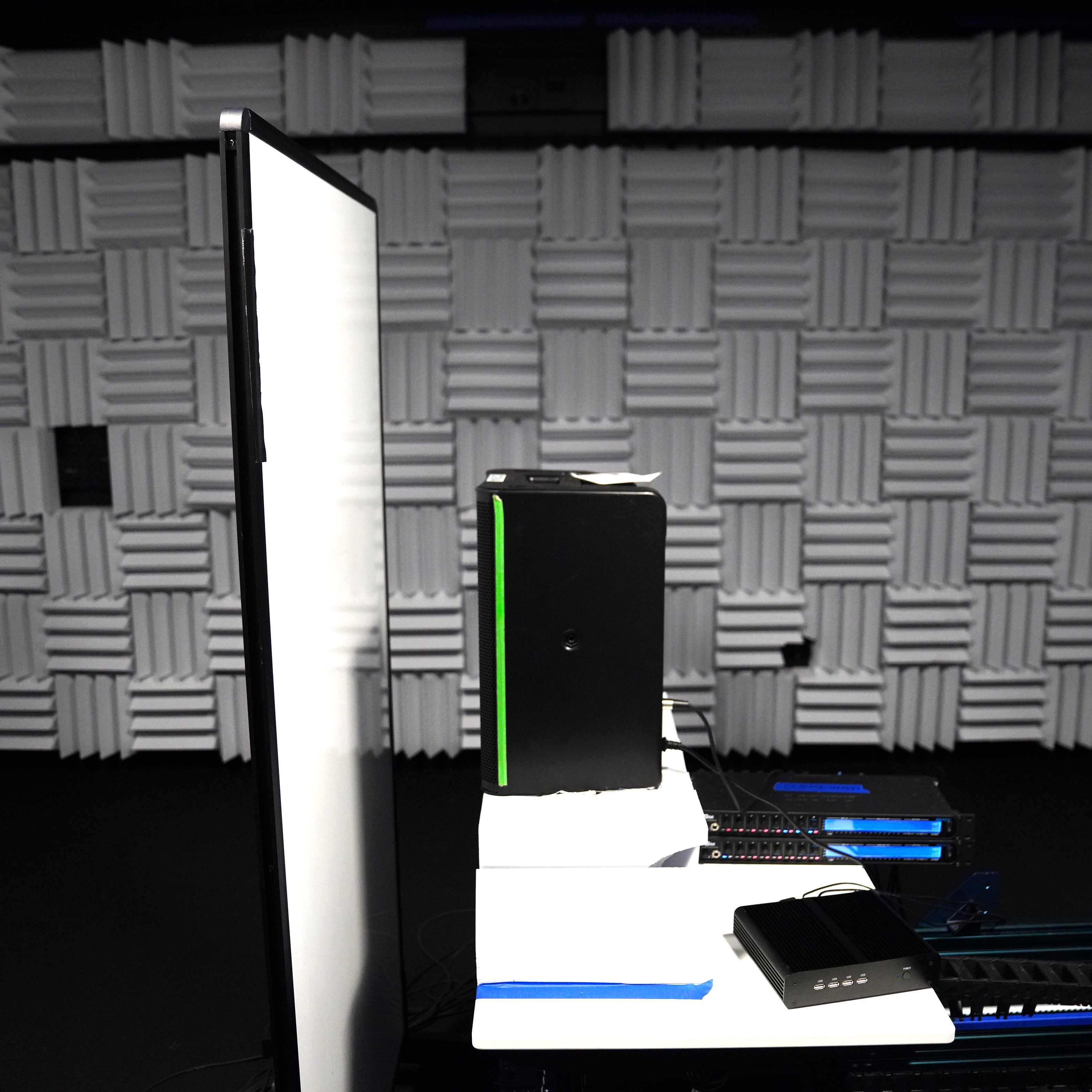} & 
        \includegraphics[width=0.18\textwidth]{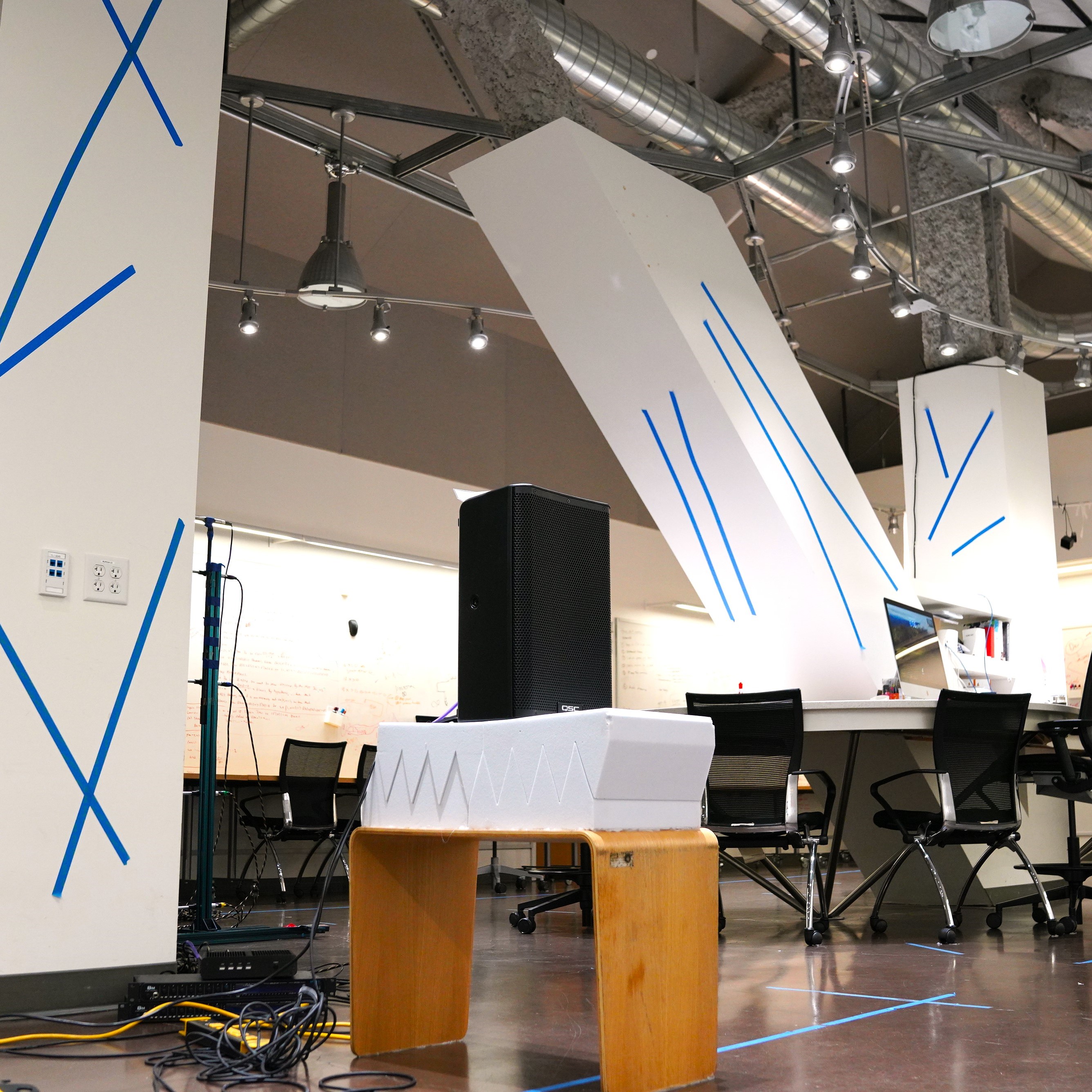} & 
        \includegraphics[width=0.18\textwidth]{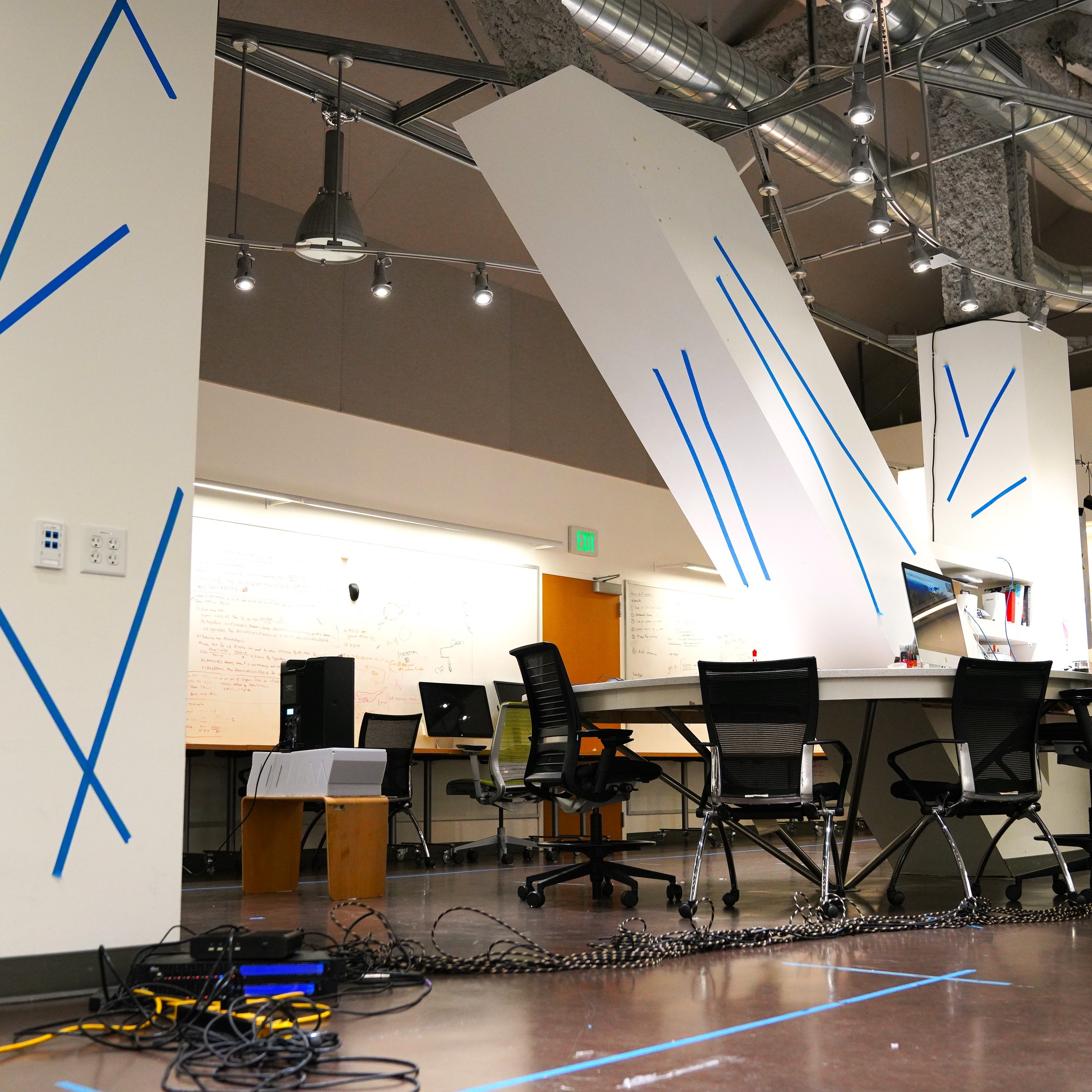} \\
        Dampened Rotated & Dampened Translated & Dampened Panel & Complex Rotated & Complex Translated \\
        \includegraphics[width=0.18\textwidth]{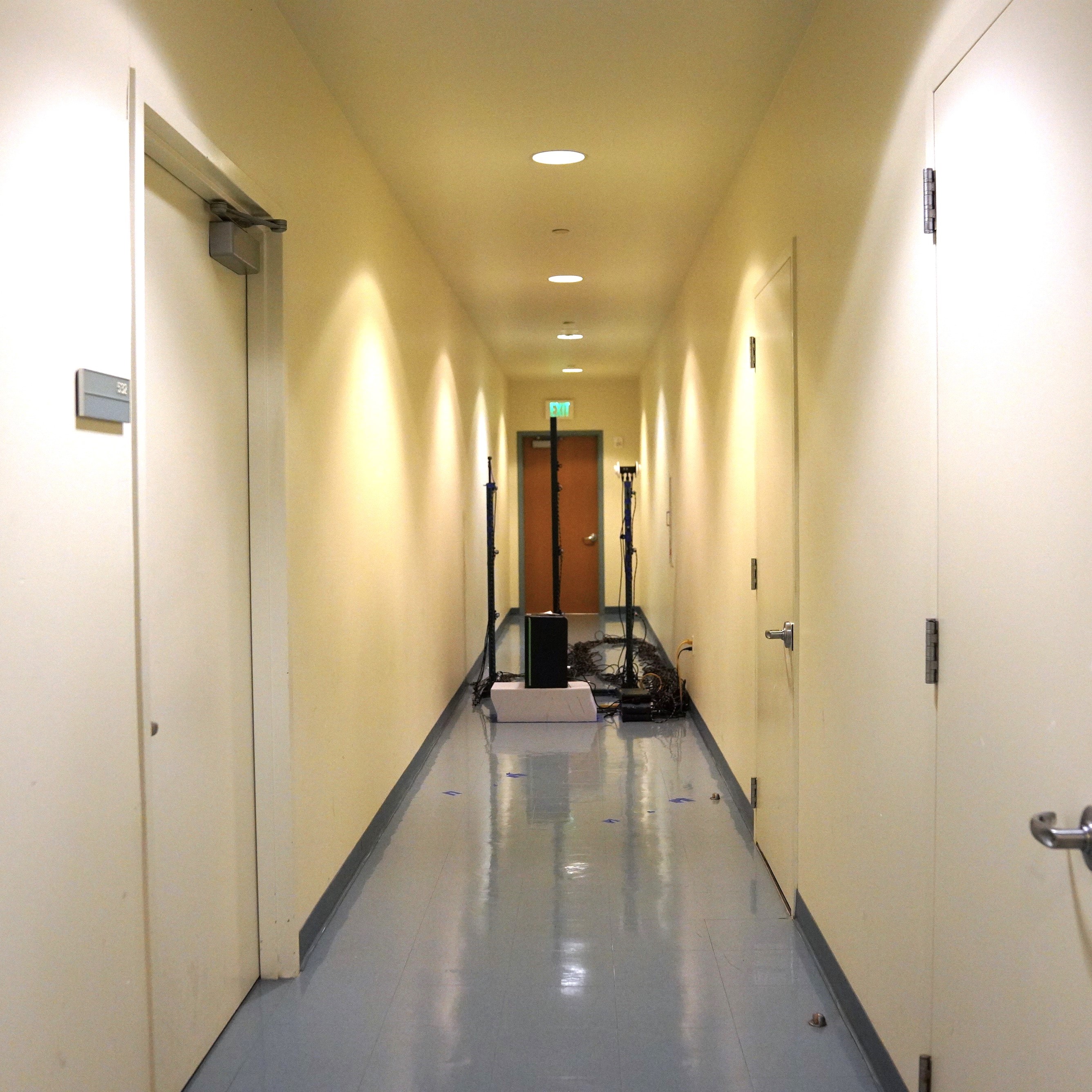} & 
        \includegraphics[width=0.18\textwidth]{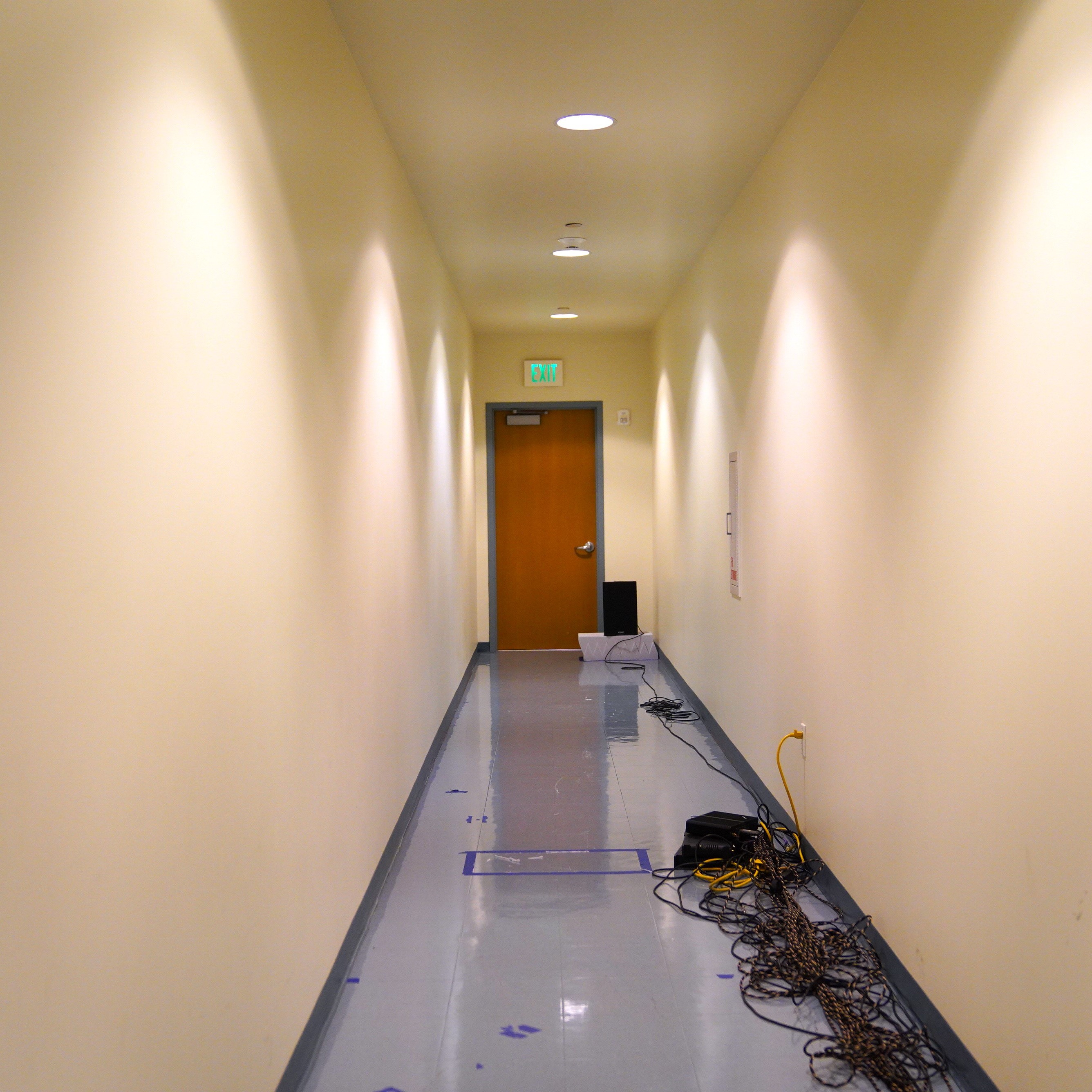} & 
        \includegraphics[width=0.18\textwidth]{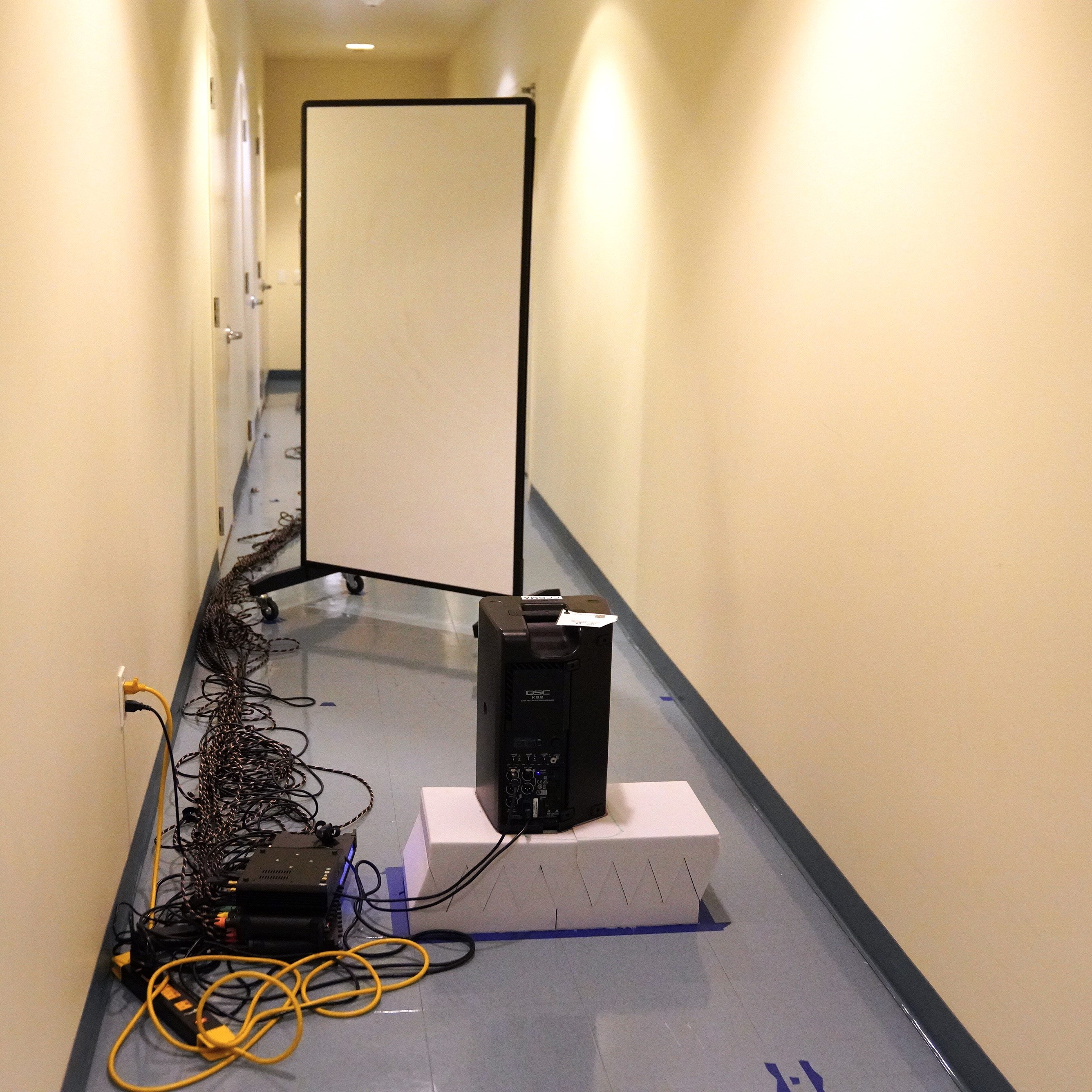} & 
        \includegraphics[width=0.18\textwidth]{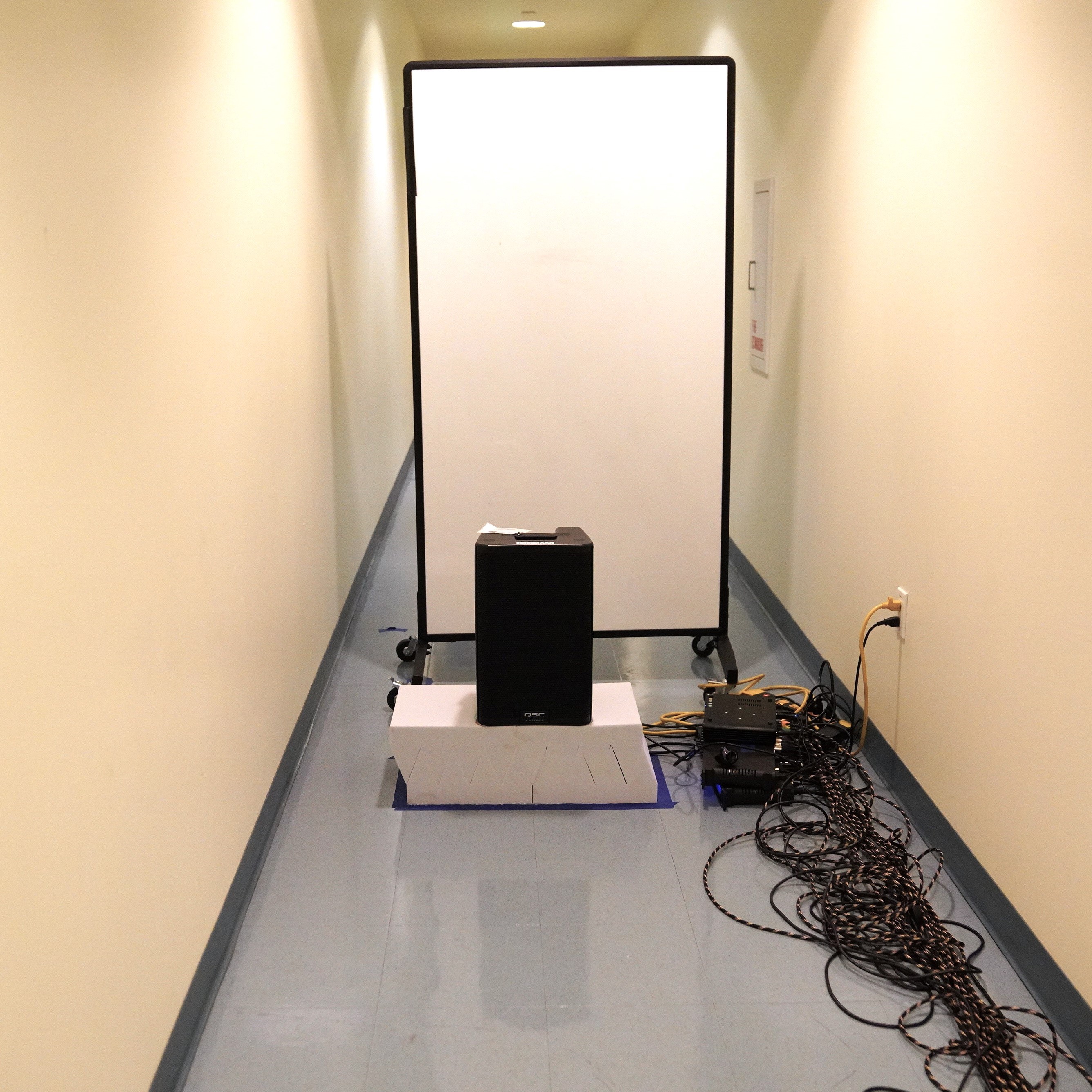} & 
        \includegraphics[width=0.18\textwidth]{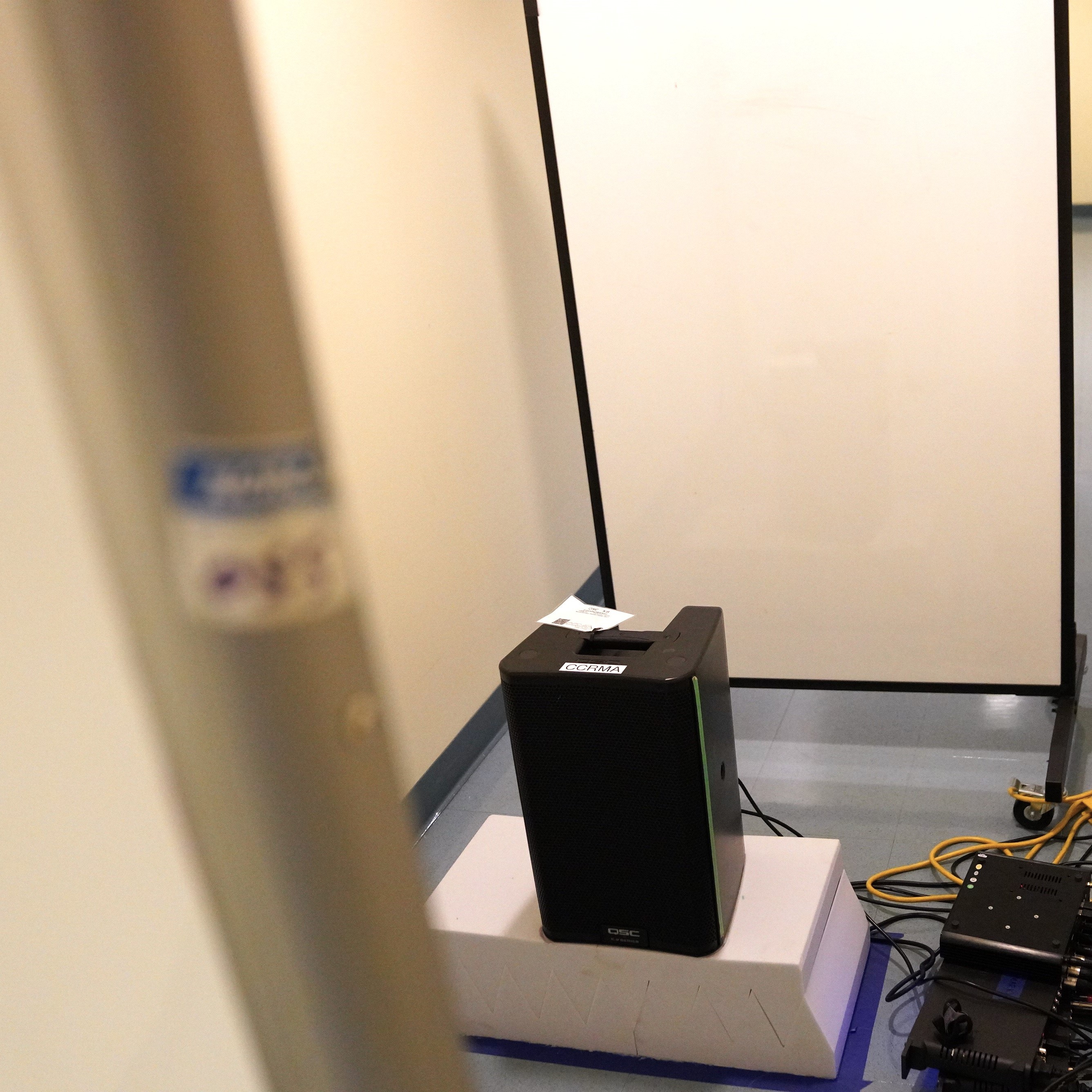} \\
        Hallway Rotated & Hallway Translated & Hallway Panel 1 & Hallway Panel 2 & Hallway Panel 3
    \end{tabular}
    \caption{Photographs of all additional configurations in the \name Dataset. Note that the Hallway Panel 1 photo is taken from behind the speaker.}
    \label{fig:additionalconfigs}
\end{figure*}

In addition to the base subdatasets collected in each of the four rooms, we collect additional data in different room configurations, where we vary the location of the speaker, the orientation of the speaker, or the presence and number of rectangular whiteboard panels. We collect this additional data for two reasons:
\begin{itemize}
    \item To test our method's effectiveness on various room layouts, including those where the speaker is occluded.
    \item To evaluate acoustic interpolation methods on the task of zero-shot generalization to changes in room layouts, by virtually simulating speaker rotation and translation, and panel relocation and insertion.
\end{itemize}

The locations and orientations of the speakers as well as the positions of the panel(s), are provided as part of the \name Dataset. Photographs of each additional configuration are shown in \Cref{fig:additionalconfigs}.

\paragraph{Rotation Subdatasets.}
In the Dampened Room, Hallway, and Complex Room, we collected 120, 72, and 132 additional datapoints where the speaker was rotated by 225\degree, 90\degree, and 90\degree clockwise, respectively. The location of the speaker and all surfaces otherwise remain the same.

\paragraph{Translation Subdatasets.}
In the Dampened Room, Hallway, and Complex Room, we collected 120, 72, and 132 additional datapoints where the speaker was translated to another part of the room, but the orientation of the speaker is was kept the same. In the Dampened Room, we move the speaker such that it is near one corner of the room and facing a wall. In the Hallway, we move the speaker to the far end of the Hallway, such that the speaker faces the entire length of the Hallway. The Complex Room is roughly divided into two halves by the table and pillars in the middle of the room. In the Complex Room, we collect additional datapoints where the speaker is translated from the one half to the other.

\paragraph{Panel Subdatasets.}
In the Dampened Room and Hallway, we place 1-2 whiteboard panels in the room. In the Dampened Panel subdataset, we place the panel directly in front of the speaker. In the Hallway subdataset, there are three panel configurations. In Hallway Panel 1, we place one whiteboard panel in front of the speaker at a slanted angle. In Hallway Panel 2, we place one whiteboard panel directly behind the speaker. In Hallway Panel 3, we place whiteboard panels both in front of and behind the speaker.

\subsection{Evaluations on Configurations}

We evaluate our model on each of these configurations independently in \Cref{tab:additionalconfigresults}, training and testing on the same subdataset. For each configuration, we select 12 training points from each of the subdatasets, and evaluate our rendered RIRs on a test set of held-out data.

\input{tables/additional-configs}

\subsection{Quantitative Results on Virtual Room Layout Modifications}

Since our model learns physically interpretable parameters for the speaker's directivity, we expect to be able to virtually simulate rotations or translations of the speaker that are unobserved in the training data. We simulate rotating the speaker by rotating the speaker's learned directivity map, and translation by moving the speaker's estimated location during path-tracing.

These predicted changes in the speaker's location or orientation can be evaluated against real data, since the \name Dataset contains additional configurations that modify the base subdataset in each room by moving or rotating the speaker. 

The quantitative results in \cref{tab:virtualRot},~\ref{tab:virtualInsertion}, ~\ref{tab:virtualPanelMovement}, and~\ref{tab:virtualTrans} show the usefulness of the \name Dataset in benchmarking the performance of methods of virtual room layout modification. Future work can use the \name Dataset to improve the performance of these tasks.
\input{tables/virtualRotation}

\input{tables/virtualInsertion}

\input{tables/virtualPanelMovement}

\input{tables/virtualTranslation}

\paragraph{Virtual Speaker Rotation. }
As an experiment, we take the \name model trained on each base subdataset with a corresponding rotated subdataset, virtually rotate the speaker by rotating the learned directivity heatmap, and predict RIRs and music at locations in each of the corresponding rotated subdatasets. We evaluate these predictions against ground-truth RIRs and music recordings from the rotated subdatasets. In addition, we compare our virtual rotation with the performance of the \name model both trained and tested on the rotated subdatasets. The results are shown in \Cref{tab:virtualRot}. Although the model both trained and tested on the rotated subdatasets outperforms our virtually-rotated model, the results are quite close in the Dampened and Complex Rooms. The results in the Hallway are worse, perhaps because the Hallway's narrow nature means that the set of direct paths from the speaker to the training locations cover a narrow range of outgoing angles.
\paragraph{Virtual Speaker Translation.} We perform a similar experiment with virtual speaker translation, evaluating against ground-truth recordings from the corresponding subdataset. The results are shown in \Cref{tab:virtualTrans}.

\paragraph{Virtual Panel Relocation.} We would like to see if we can learn the reflective characteristics of a surface in one room, then `virtually move' the surface to another location in the same room. In the Hallway, we collect two subdatasets (Hallway Panel 1 and Hallway Panel 2 in \Cref{fig:additionalconfigs}), where the room layouts are identical except for the location and orientation of a single whiteboard panel. In our experiments, we train on the first panel configuration, then move the location of the whiteboard panel to that of the second configuration before performing inference. We then evaluate our predicted audio against ground-truth audio from the second configuration. Results are shown in \Cref{tab:virtualPanelMovement}. The baseline shown is one where we train on the same subdataset that we evaluate on.

\paragraph{Virtual Panel Insertion.} We would like to see if we can learn the reflective characteristics of a surface in one room, then `virtually insert' the surface into another room.
Three of our base subdatasets also include a version with a single inserted whiteboard panel. In each of our four experiments, we take the base subdataset (\eg the Dampened Base subdataset), and the coefficients learned for the whiteboard panel from another room (\eg the Hallway Panel Config. 1 subdataset). We then virtually insert the whiteboard panel into the base subdataset, and evaluate the virtual insertion against the version of the base dataset with a panel in it (\eg the Dampened Panel subdataset). Results are shown in \Cref{tab:virtualInsertion}. The baseline shown is one where we train on the same subdataset that we evaluate on.

% \paragraph{Virtual Panel Insertion.} Since our framework also attempts to learn the reflective characteristics of each surface, we can also attempt to virtually insert surfaces into scenes in a zero-shot fashion. These surfaces can have predefined reflection coefficients, or reflection coefficients that have been discovered by training the model in another room. In practice, this does not work, perhaps because it is difficult to cleanly disentangle the sound source's response, the directivity heatmaps, and the reflective properties of the surfaces in the room. However, the \name dataset provides these configurations to aid the development of methods to simulate such zero-shot surface insertions.

\section{Additional Experiments and Ablations}

\subsection{Results on Binaural Rendering}
We evaluate our method on the task of rendering a binaural RIR at an unseen location. We collect binaural RIRs at several locations in all rooms using our 3Dio binaural microphone, and compare these to predicted RIRs that we binauralize from single-channel audio as described in the Methods section.

We compare our binauralized audio with the ground-truth audio using the left-right energy ratio error between the ground-truth and predicted recordings, which is used in~\cite{chen2023novel}. To compute the left-right energy ratio, we compute compute the ratio of total energy between the left and right channels of the RIR or music recordings. We then compute the mean squared error between the left-right energy ratio of the predicted and ground-truth RIRs or music. Results are shown in \Cref{tab:binaural_results}.

Since the baselines do not have a way of generating binaural RIRs from monaural ones, we binauralize these baselines by rendering two monaural RIRs at the locations of the left and right ears of the 3Dio microphone, and combining them into left and right channels.
% \begin{equation}
%     \left(\frac{\hat{E}_L}{\hat{E}_R} -  \frac{E_L}{E_R}\right)^2
% \end{equation}\label{eq:LRE}

% Where $E_{L,R}$ is the total energy in left or right channel of the RIR or music spectrogram

Our method outperforms our baselines across most metrics. Note that it is difficult to compare a binaural RIR recorded from our binaural microphone with binauralized audio originally recorded from a different microphone. Our rendered binaural audio will have characteristics of the monaural microphone and the microphones used in the SADIE dataset~\cite{armstrong2018perceptual} used to record the HRIRs that we convolve our monaural recordings with. The binaural recordings in our dataset will h    ave different characteristics, since they are recorded using a different microphone with different spectral characteristics and directionality. Because of this, we include qualitative binauralization results in the supplementary video.

% Mason - Check Wording

\input{tables/bin}

\subsection{Performance vs Number of Training Points}

\begin{figure}[!t]
    \centering
    \includegraphics[width=\columnwidth]{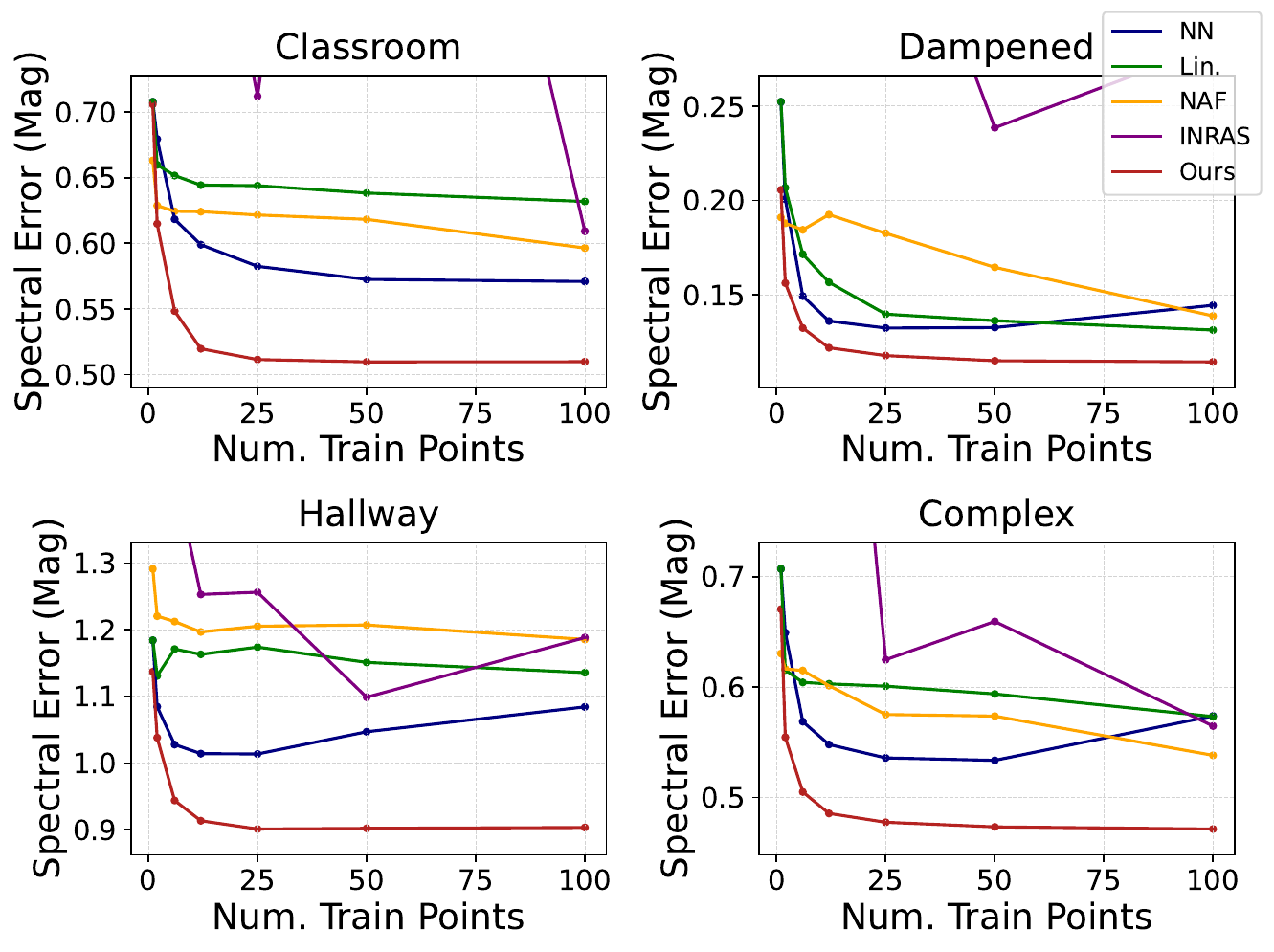}
    \vspace{-20pt}
    \caption{Evaluations of our method and baselines with different numbers of training points. We use the Multiscale Log-Spectral L1 Loss (Mag), and train with $N\in\{1,2,6,12,25,100\}$. All training locations are selected as nested subsets of one another, and we evaluate on a fixed test set. Note that the DeepIR baseline's error was too large to fit into the range of the plot.}
    \label{fig:num_train_exp}
\end{figure}

We conduct an ablation study with varying numbers of training points $N$ on each subdataset and compare against the baselines.
As shown in \Cref{fig:num_train_exp}, the performance increases with $N$, and our model consistently outperforms the baselines when $N \ge 2$. Note that in all rooms, our model trained on only 6 locations outperforms all baselines trained on 100.

Note that our model's hyperparameters are optimized for performance in data-limited scenarios. When the number of training points is higher, it is possible that increasing the number of parameters (for instance, increasing the resolution of the heatmap or the number of reflection coefficients) leads to even better performance.

\subsection{Robustness to Inaccurate Geometry.}
\begin{figure}[!t]
    \centering
    \includegraphics[width=\columnwidth]{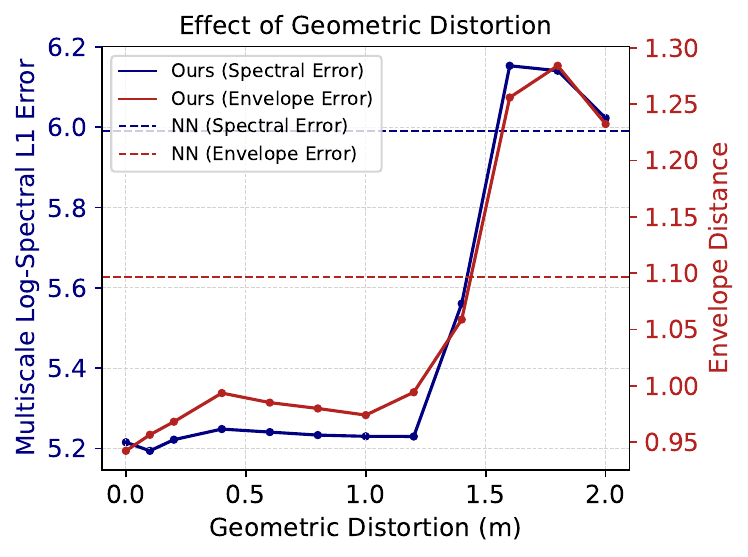}
    \vspace{-20pt}
    \caption{Effect of geometric distortion on RIR prediction performance in the Classroom subdataset The blue line shows our model's performance according to the Multiscale Log-Spectral L1 metric, and the red line shows our model's perfomance according to the envelope distance metric. The red and blue dashed lines indicate the performance of the nearest-neighbors baseline according to the multiscale log-spectral L1 metric and the envelope distance metric, respectively.
    }
    \label{fig:warp_exp}
\end{figure}

Our method requires measuring the room's geometry. In our dataset, we do this using a tape measure or laser distance measure, which both provide sufficiently accurate measurements. In order to explore the effect of inaccurate geometric measurements, we conduct an additional experiment to measure the performance after adding random artificial distortions to the surfaces. In the Classroom, we select 8 random directions to move each of the 11 vertices defining the walls, ceiling, floors and the corners of the tables that are exposed. We move each vertex by 0-2 meters in its corresponding random direction. Results are shown in \Cref{fig:warp_exp}.  Observe that unless we distort \emph{all} vertices in the room by over 1.5 meters, our model outperforms the best baseline (Nearest Neighbors). We conclude that our method is robust to geometric distortion.

Geometric distortion can affect our model's rendering in one of three ways: It can change the distance of reflection paths, which affects its time-of-arrival and amplitude; it can eliminate reflection paths, or it can add new reflection paths. Since our model is optimized against a frequency-domain loss whose smallest window size is 256 samples (or 1.8 meters at the speed of sound), our model is robust to perturbations in times-of-arrival.

\camerareadysection{
\subsection{More Ablations}
In the Methods section, \Cref{sec:boosts}, and \Cref{sec:hopsize1}, we discuss several minor components of our model (axial boosting, time-of-arrival perturbation, hop size 1 loss, etc.) that provide a boost to our model's performance and/or robustness. Results with each of these components ablated are in \Cref{tab:ablations2}. Our model performs the best on a plurality of evaluations, proving that these performance boosts are good on balance. However, we should also observe that even in evaluations where our model does not perform the best, it is never worse than the best performing ablation by a significant margin. We cannot say the same for the Interpolation Spline ablation, which also performs the best in the same number of evaluations (six), but significantly underperforms our model in several settings.

\input{tables/more-ablations}

\subsection{Modeling the Effects of Transmissions}

Our model assumes that sound energy encountering a surface is either reflected or absorbed by the surface. This is for the sake of simplicity. We also conduct an experiment in which we consider surface transmission as well. This means that we modify our tracing algorithm to consider reflection paths that pass through surfaces, and assume that a proportion of the sound energy at each frequency can be \textit{transmitted} through these surfaces in a frequency-dependent manner. Our modified training procedure then fits \emph{surface transmission coefficients} in a manner identical to the way it fits \emph{surface reflection coefficients}. \Cref{tab:transmission} contains quantitative results from a model that models both transmission and reflection, and shows that in our settings, modeling transmission is not necessary. However, in other rooms with surfaces of different materials, modeling transmission may be important.

\input{tables/transmission}

\subsection{Comparison to Traditional Acoustic Simulations}

We compare our method to a widely-used image-source audio simulator, Pyroomacoustics~\cite{Scheibler_2018}. For each room in our dataset, we simulate RIRs by providing the dimensions and the speaker location, and selecting the closest material coefficients for each surface from its pre-defined database (\eg drywall, ceiling tiles, carpet).
\Cref{tab:pyroom} reports the accuracy of the simulated RIRs compared to the ground truth.

\input{tables/pyroom}

}

\section{Method Details}
\label{app:method}
\subsection{Details on Source Localization}\label{app:toa-method}
\camerareadysection{Our method does not require a ground-truth source location measurement}. Instead, we use a simple time-of-arrival \camerareadysection{technique} to estimate the sound source's location to a degree of accuracy sufficient for the subsequent steps of the method. For each Room Impulse Response (RIR) in the training set, we determine its first peak, which is proportional to the distance of the direct path between the microphone and source locations. We locate the first peak of the RIR by measuring when the RIR first exceeds a quarter of its absolute maximum. We then determine the distance from the source to the target microphone by multiplying by the speed of sound, assumed to be 343 m/s.

We use a gradient descent optimization method to fit the optimal source location. We initialize the source location to the origin, which is at a corner of the room. We perform an optimization process that updates the optimal source location's position at each step. At each iteration of the optimization process, we compute the estimated times-of-arrival for each of the microphone locations, based on the current estimate for the source location. We then calculate the L1 loss between the estimated times-of-arrival and the times-of-arrival as measured by locating the first peak of the ground-truth RIR as described in the previous paragraph. We perform a gradient update on the estimated source location to minimize this L1 loss. We optimize for 1000 steps, and use the final estimate for the source location as our estimated source location.

\camerareadysection{
In all of the base configurations, our method is able to predict a source location that is inside the location of our QSC loudspeaker. We used the estimated location in all configurations except for the Complex Rotation and Complex Translation configurations, where our localization method failed.}

% !@#$ Mason - could 

\subsection{Minimum-Phase Transform}\label{sec:minphasejust}
Our model learns the frequency-domain response curve for each of the surfaces in the room and for each outgoing direction from the source, allowing us to determine how the frequency profile of sound traveling along that reflection path is altered. However, this frequency-domain response is not enough to determine the reflection path's time-domain contribution to the RIR, because it contains magnitude information, but no phase information. In order to invert our reflection path's frequency profile into a time-domain signal, we need to provide phase values at each frequency, so we can perform the inverse-Fourier transform.

\camerareadysection{In our analysis, we adopt the minimum-phase assumption to calculate phase values for acoustic reflections, a method widely recognized and justified within acoustic research~\cite{minphasejustification, minphasejustification2}. This assumption posits that for each frequency, the phase delay introduced by the reflection is minimal, implying that the time delay contributed by the path of reflection at any given frequency is as short as possible. From a physical standpoint, this is akin to assuming that sound is reflected off surfaces with negligible delay, thereby behaving as if the reflections are `instantaneous' while still preserving the unique frequency-dependent characteristics of the reflection.}
We compute the phase values using the method described in~\cite{SASPWEB2011}.

\subsection{Specific Loss Formulation}\label{sec:lossformula}

\camerareadysection{

\paragraph{Loss Formulation and Equations.}
We define the loss for a given short-time Fourier transform (STFT) window size $s_w$ and hop size $h$ in \Cref{eq:one_size_loss}. This is the sum of the L1 distance between the magnitude-spectrograms  of the ground-truth and synthesized RIRs and the log-magnitude spectrograms of the ground-truth and synthesized RIRs.

\begin{figure*}[!h]
\begin{equation}\label{eq:one_size_loss}
L_{s_w, h}(W, \hat{W}) = |S(W, s_w, h) - S(\hat{W}, s_w, h)| + 
|\log S(W, s_w, h) - \log S(\hat{W}, s_w, h)|
\end{equation}
\begin{equation}\label{eq:total_loss}
L(W, \hat{W}) = \left[\sum_{s_w \in (512,...4096)}  L_{s_w, Hs_w}(W, \hat{W}) \right] + L_{256, 1}(W, \hat{W})
\end{equation}
\end{figure*}

In the formula, $W$ and $\hat{W}$ indicate the ground-truth and predicted RIRs, respectively. $h$ indicates the hop length, $s_w$ indicates the  STFT window size, and $S$ is the short-time Fourier transform, or spectrogram, whose arguments are the time-domain signal to transform, the window size, and the hop length, respectively. $H$ indicates the hop ratio, or the hop length divided by the window size. We set $H=0.25$.

\Cref{eq:total_loss} provides the total loss, which sums across multiple window sizes, and adds a loss term that uses a hop size of 1. 

\paragraph{Hop Size 1 Loss. }\label{sec:hopsize1}
We use a spectral loss term with hop size 1 to ensure that the early part of the RIR has accurate time-domain characteristics, since the hop length of 1 allows for high-resolution in the time domain. We take inspiration from~\cite{chen2022sound} for this term, and discover it improves performance, as seen in \Cref{tab:ablations2}.

\paragraph{Modifications from Related Work.}

Our multi-scale \camerareadysection{spectral plus} log-spectral loss is identical to those used in~\cite{clarke2022diffimpact} and \cite{engel2020ddsp}, with two exceptions: First, is the introduction of the loss term with hop size one. Second, the minimum window size in our loss is $256$, instead of $32$ or $64$. This is because there will be error in the time-of-arrival of certain reflection paths, due to geometric measurement error (which increases with reflection order) or errors in the speed of sound approximation. This means that the placement in time of a reflection path's contribution to our synthesized RIR may be off from its placement in the ground-truth RIR by some amount. Using larger window sizes compensates for this error, since larger windows are more likely to contain both the reflection path's contribution to our synthesized RIR and its contribution to the ground-truth RIR.
}

\subsection{Small Efficiency and Performance Boosts}\label{sec:boosts}

\paragraph{Efficiency Boosts.} Since each rendered RIR combines hundreds of reflection paths, we compute all the reflection path contributions in parallel to minimize runtime. In addition, all reflection paths for the training points are precomputed before training starts.

\paragraph{Time-of-Arrival Perturbation.} Since our measurements of each room are not necessarily precise, to make our model more robust, especially with an extremely limited number of measurements, we would like to perturb the surfaces during training. However, reflection paths for all training locations are precomputed before the training process begins. Perturbing each surface would require retracing at each iteration, which is computationally inefficient. As a proxy to this, we perturb the time of arrival of all paths by adding Gaussian noise to it, with a standard deviation of 7 samples. We found that this improved the interpretability of the estimated parameters and led to minor perfomance boosts, as shown in \Cref{tab:ablations2}.

\paragraph{Regularization via Convolution with Pink Noise.}

Since RIRs are often used as a means to simulate sounds in an acoustic environment, we would like to not only ensure that our rendered RIRs are accurate, but also that the sounds we simulate via convolution with the RIR are accurate. Minimizing the spectral loss between ground-truth and predicted RIRs does not always accomplish this, since convolving the RIRs with other waveforms results in significant changes in the spectrograms. 

Pink Noise is a special type of noise whose power spectral density is inversely proportional to frequency. It is ubiquitious in nature~\cite{pinknoise}, and is often used as a test signal to calibrate sound systems and loudspeakers, since its frequency profile is similar to that of music~\cite{Davis1987SoundReinforcement} and other sounds the speaker might play.

To encourage our model to maintain accuracy post-convolution, we implement a regularization strategy using pink noise. For the latter half of training iterations, we convolve both our predicted and the ground-truth RIRs with five seconds of randomly generated pink noise, compute the loss between them, and add it to the loss computed between RIRs at each iteration. Convolving RIRs with pink noise simulates the speaker playing of a pink noise test signal. It is equivalent to reshaping the RIR's spectrum according to the profile of pink noise, and applying a random phase shift at each frequency. 

\Cref{tab:pinkconvolve} shows that this form of regularization results in improvements in both RIR prediction and music prediction. Such forms of regularization should be the study of future work and theoretical study.

With the goal of improving rendered music in mind, we also tried a similar form of regularization, where we convolve both our ground-truth and predicted RIRs with five seconds of music randomly sampled from the FMA dataset~\citep{defferrard2016fma} at each iteration after training is halfway done. Convolution with the music files simulates the speaker playing them. Results for this form of regularization are also shown in \Cref{tab:pinkconvolve}, although we prefer the performance and simplicity of pink noise regularization.

\input{tables/pink}

\subsection{Computational Cost}
\paragraph{Training and Path-Tracing Time.}
In all of our experiments, we trained our model for 1000 epochs. In \Cref{tab:times}, we report the amount of time it took for our model to train on each of the base room configurations. Note that since the Complex Room is only traced up to order 4, there are substantially fewer valid reflection paths, and thus training is faster. \camerareadysection{In all other rooms, we trace up to order 5. Tracing is slower in rooms with more surfaces.}

\paragraph{Main Contributions to Training Time.} \camerareadysection{We also measured the different steps in the training process to see which ones took the longest. Each training location is associated with hundreds of reflection paths that must be added together to form the the RIR. While rendering these contributions is done in parallel, compiling them requires placing them in at the right locations in time and is done sequentially. In practice, 37.7\% of the time during the 1000 epochs is spent on this compilation, 61.9\% on the backwards passes, and 0.4\% on everything else.}

\begin{table*}
\small
\centering
\begin{tabular}{lccccc}
\toprule
Room & Training Time (Hours) & Inference Time (s) & N. Surfaces  & Tracing Time (s) & Avg N. Reflection Paths\\
\midrule
Classroom & $9.61$ & 0.90 & 9 & 4.3 & 874
\\
Dampened & $5.75$ & 0.56 & 6 & 0.83 & 675 
\\
Hallway & $8.97$ &  0.90 & 6 & 1.5 & 853
\\
Complex &  $2.82$ & 0.37 & 33 & 47 & 439
\\
\bottomrule
\end{tabular}
\caption{In all of our experiments, we train our model for 1000 epochs and report the training time that this takes in each room, in the base configuration. In addition, we report the inference time, or the time it takes our model to render a single RIR. Before training begins, we precompute the reflection paths that go between the source and listener locations, up to a certain maximum reflection order, so we also report this tracing time to trace reflection paths, per listener location of each room and its corresponding subdataset. We also report the number of valid reflection paths found by the tracing algorithm, as an average across all points in the subdataset.
}
\label{tab:times}
\end{table*}

\section{Baseline Implementation Details}
 \paragraph{Linear.} The Linear baseline computes a RIR at a given test location by taking a linear combination of the four nearset points in the training data. The weights on each of these four training points are inversely proportional to the distance to the test location. We also experimented with taking a weighted combination of all the training data, where the weights are inversely proportional to distance. This alternative linear baseline performs quite poorly, with error increasing with the number of training points. This is because the training RIRs are roughly uncorrelated with mean zero, so the average of $N$ RIRs tends towards zero as $N$ increases. 
 \paragraph{Neural Acoustic Fields (NAF) \cite{luo2022learning}.}\:\:
        To compare our method to NAF, we utilized NAF's official code,\footnote{\url{https://github.com/aluo-x/Learning_Neural_Acoustic_Fields}} as open-sourced by authors. 
        However, in order to apply NAF to our dataset and experimental settings, we modified this code in some minor ways.
        Specifically, the original NAF was designed to estimate arbitrary stereo RIRs constrained to lie on a 2D horizontal plane within a 3D room, \ie it did not consider a $z$-axis and thus does not output RIRs at arbitrary heights. 
        Therefore, we added the height on the $z$-axis as an input, embedding it by using the same positional encoding~\cite{pmlr-v97-rahaman19a, mildenhall2021nerf} as the authors’ code. The corresponding elements of the network architecture, \eg the number of units in the input layer, were also modified. 
        The architecture we used for NAF in our experiments consisted of 8 linear layers with Leaky-ReLU activations~\cite{maas2013rectifier}. Note that we only changed the number of the number of units in the input layer, from 126 to 168, due to the aforementioned addition of $z$-axis features. 
        In addition, the NAF we used in our experiments was designed to output only magnitude-spectrograms, \ie without any phase information, because the official code also does not have the phase-related loss and corresponding phase output.
        We utilized the Griffin-Lim algorithm \cite{griffin1164317} to estimate the phase of each magnitude spectrogram and render the time-domain RIRs.
        For training, we followed the same process in their official code and used the model's weights after the final training epoch for inference and evaluation. Finally, we used a 48000 Hz sample rate rather than the original 22050 Hz.
        All other settings, such as the optimizer, number of epochs, learning rate, etc., are the same as their official implementation.
        \paragraph{Deep Impulse Responses (DeepIR) \cite{richard2022deep}.}\:\:
        Unlike NAF, the authors of DeepIR have not open-sourced an official codebase. 
        Therefore, we implemented DeepIR ourselves, based on the details in their paper. 
        Specifically, we built a simple multi-layer perceptron (MLP) consisting of 6 linear layers, each followed by leaky-ReLU activations.
        The input feature vector consists of $(x, y, z, t)$, which are the desired spatial coordinates and the time index, respectively.
        Similar to NAF, we applied positional encoding to all inputs before passing them into the MLP.
        Hence, the number of units in the input layer is $d_{\rm{emb}}$, whereas all other layers have 512 units.
        DeepIR directly outputs the $t^{\text{th}}$ time sample of the RIR to produce an estimate $\hat{\mbox{IR}}$ of the full RIR. We then convolve this with the arbitrary dry source audio $x$, to produce an estimate $\hat{y}$ of the sound of the arbitrary audio being recorded from the specified source and listener location in the room, \ie $\hat{y} = x * \hat{\mbox{IR}}$. We optimized $\hat{y}$ according to an L2 loss comparing the log-magnitude spectrogram with that of the corresponding ground-truth audio $y_{\rm{gt}}$.
        We omitted the noise model, since our dataset did not include artificially added noise, and the noise in our recordings was minimal.
        We set other hyperparameters for DeepIR such as the optimizer, learning rate, the number of epochs, etc., to the same values as NAF.
\paragraph{Implicit Neural Representation for Audio Scenes (INRAS)~\cite{su2022inras}.}
The authors of the INRAS baseline released their code in the Supplementary Materials of their submission.\footnote{\url{https://openreview.net/forum?id=7KBzV5IL7W}} We use their code with some minor modifications. The framework is originally trained and tested on data from the SoundSpaces dataset~\cite{chen2020soundspaces}, which provides simulated binaural recordings within virtual environments. The architecture is built around consuming this data, where each simulated recording represents a stereo, binaural recording with the head positioned at one of the four cardinal angles. Our training sets use exclusively monaural recordings from omnidirectional microphones. Thus, in order to make our changes to the network as minimal as possible, we duplicated our mono-channel recordings to stereo-channel recordings and assumed them to all be at the $0^{\circ}$ angle. We then took only the left channel of the stereo output as the framework's estimate of the monaural RIR. Since INRAS consumes environment meshes, we provide it with a 3D scan of each room. Otherwise, we used mostly the same hyperparameters as the original, with the exception that we increased the sample rate from 22050 to 48000 Hz. Since our training set of 12 recordings per subdataset was approximately four orders of magnitude smaller than the datasets on which the authors had trained, we increased the initial learning rate from to 0.001 instead of 0.0005, slowed the learning rate's exponential decay schedule to decay rate $\gamma=0.1$ over 3000 epochs rather than 50, and trained for 5000 epochs rather than 100. We evaluated the model against a validation set every 100 epochs. For our test evaluations of INRAS, we used the weights and consequent outputs of the model with the best performance across all such validation evaluations.

\section{Data Collection Procedure Details}
\begin{figure}
    \centering
    \includegraphics[width=\columnwidth]{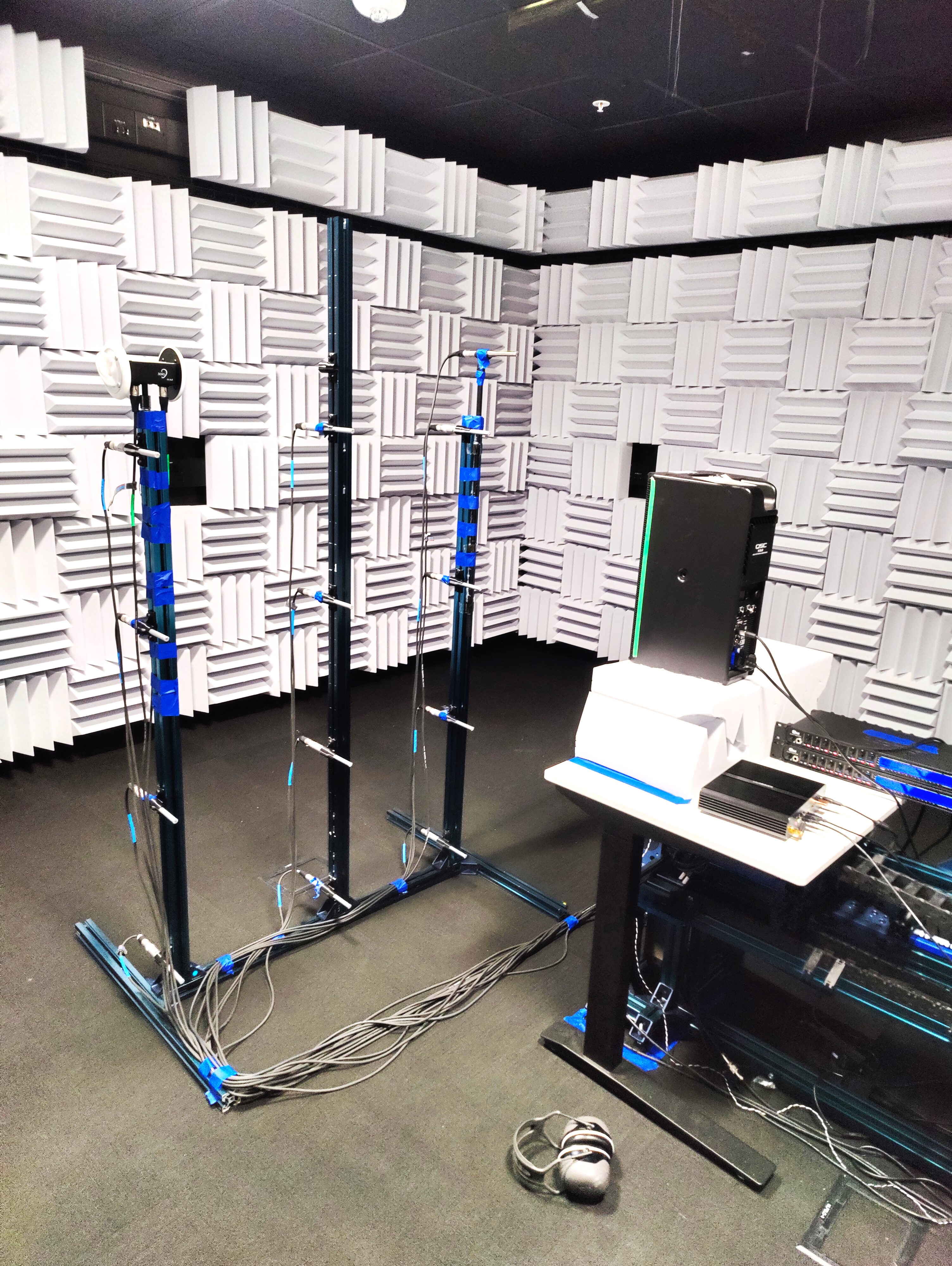}
    \caption{A photo of the data collection procedure in the Dampened Room. The custom microphone frame holds 12 microphones, as well as a 3Dio FS XLR binaural microphone.}
    \label{fig:micframe}
\end{figure}

We use a custom-built microphone frame designed to accommodate 12 Dayton Audio EMM6 measurement microphones, as well as one 3Dio FS XLR binaural microphone, all of which were rigidly mounted at precisely measured positions on the frame. \Cref{fig:micframe} shows a photo of the microphone frame used to collect the data. We set the origin of each room such that there is one wall representing $x=0$ and one wall representing $y=0$. Before each recording, we positioned the frame within the room and measured the distance from the edge of the frame to each of the origin walls using a tape measure or a Bosch GLM20 laser distance measure, which have 1 and 3 millimeter measurement resolutions, respectively. We use the measured position of the frame's corner as well as the pre-measured offset of each microphone from the frame's corner in order to annotate each microphone's position in the room to sub-centimeter precision for our dataset.

\subsection{Estimating the Room Impulse Response (RIR)}\label{app:measure_rir}

In order to measure each RIR, we played a logarithmic sine sweep through the speaker. The sweep spanned from 20 Hz to 24 kHz for 10 seconds, followed by 4 seconds of silence. This sine sweep was recorded from each of the microphones simultaneously at each gantry position. While sending the sine sweep signal from the audio interface to the speaker, we also recorded loopback signal by wiring the audio interface's output to one of its inputs. We used this loopback signal to estimate and correct for the latency in the system.

To compute the RIR $r[t]$, we take
$$
r[t] = IFFT \left( \frac{FFT(a[t])}{FFT(l[t])} \right),
$$
where $FFT$ and $IFFT$ are the Fast-Fourier Transform and its inverse respectively, $a[t]$ is the digital recording of the sine sweep, and $l[t]$ is the digital loopback signal. Note that we deconvolve the loopback signal from the recording, instead of deconvolving the source signal sent to the speaker from the recording. We assume that the loopback signal is the same as the source signal, but delayed in time by the latency of the system. Deconvolving from a delayed copy of the source signal instead of directly from the source signal thus corrects for the delay in the system. We remove the last 0.1 seconds of the 14-second RIR to eliminate anti-causal artifacts.

\camerareadysection{In addition, to account for differences in microphone sensitivity, we adjust the volume of each sweep recording according to the sensitivity of the microphone used to record it. Specifically, we look up each EMM6's microphone's response at 1000 Hz in dB from its calibration sheet, and reduce the overall volume of its recordings by the same amount.}

\subsection{Room Geometry Estimation}
As the wavelengths of audible sound typically range from 2 cm - 17 m~\cite{moller2004hearing}, the prominent sound waves are likely to bypass or diffract around smaller surfaces.
Hence, we only focus on modeling salient surfaces (\eg walls, pillars, table tops), which are often characterized by planes, and simply trace the reflection paths using image source methods.
For the rooms we captured in our dataset, we also measured the walls and surfaces and manually created planar mesh-based reconstructions of them.
With the recent progress in visual 3D scene reconstructions~\cite{mildenhall2021nerf}, our geometric estimation can also easily be replaced by automatic algorithms or even mature customer tools such as Polycam.

\section{Guidelines for Microphone Placement.}
\begin{figure}
    \centering
    \includegraphics[width=0.75\columnwidth]{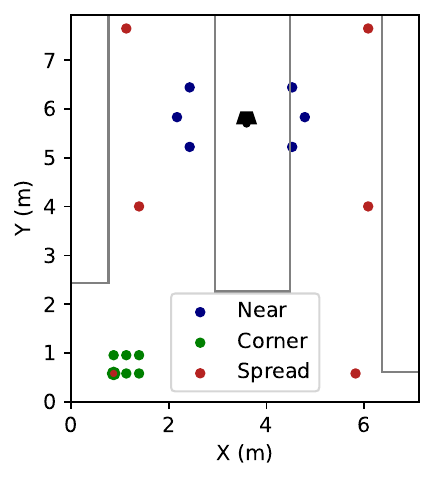}
    \caption{The distributions of three different sets of training points in the Classroom subdataset. The grey lines indicate the locations of tables in the subdataset.}
    \label{fig:trainingdist}
\end{figure}

\input{tables/mic_placement}
To maximize efficiency, we found it empirically beneficial to spread our RIR locations in all three dimensions. This allows us to 1) cover a variety of angles around the speaker, which likely leads to better speaker directivity estimates, 2) disentangle the effects of individual reflections, and 3) better estimate the diffuse sound field, which is approximated in our model as spatially uniform.

To study this effect, we conducted an experiment in the Classroom subdataset. We select three different sets of training locations (shown in \Cref{fig:trainingdist}), each of which contain 6 RIR recordings from 6 different locations. For simplicity, these training locations were selected in the 2D plane defined by $Z=0.98$. We evaluated \name trained on each of these sets of training locations on a test set comprised of other points selected in the $Z=0.98$ plane.

Our best performance across all metrics is achieved in the `Spread' configuration of training points, confirming our intuition. Interestingly enough, the `Near' Configuration performed the worst. We believe this could be due to the model overfitting to the near-field of the speaker~\cite{Howe_2002nearfield}, which can be substantially different than the sound field at other locations in the room.

% \section{Downstream Applications}

% Note that in addition to being able to estimate the audio at novel viewpoints, our method extracts interpretable parameters for important intrinsic properties of the acoustic environment. These parameters not only provide us with interesting insights into the acoustic properties of the surfaces and sound source, they also can be reused for interesting downstream tasks. Using these parameters, we may even modify the position or orientation of the sound source, or modify the configuration of surfaces in the room, then without any additional training data from the modified environment, estimate the acoustics at novel viewpoints within the modified environment. This task is not well-defined for any of our baselines, so evaluating them quantitatively against our method unfortunately is not feasible. However, future work could focus on making this framework composable, so that different objects with different learned characteristics can be virtually placed in new scenes.

%% file: tables/additional-configs.tex
\begin{table}
\small
\centering
\begin{tabular}{lccccc}
\toprule
 & \multicolumn{2}{c}{RIR} &  \multicolumn{2}{c}{Music} \\
 \cmidrule(lr){2-3} \cmidrule(lr){4-5}
\textbf{Room/Configuration} & Mag & ENV & Mag & ENV \\
\midrule
Dampened & 1.21 & 0.56 & 1.59 & 1.19\\
\:\:\:w/ Rotated Speaker & 1.14 & 0.44 & 1.49 & 1.36\\
\:\:\:w/ Translated Speaker & 0.68 & 0.39 & 0.91 & 1.18\\
\:\:\:w/ Panel & 1.23 & 0.60 & 1.62 & 1.47\\
\cmidrule(l{1em}r{1em}){1-5}
Hallway & 9.13 & 2.95 & 2.59 & 1.25\\
\:\:\:w/ Rotated Speaker & 8.40 & 2.86 & 2.58 & 1.27\\
\:\:\:w/ Translated Speaker & 8.91 & 3.02 & 2.84 & 1.25\\
\:\:\:Panel Config. 1 & 8.47 & 2.99 & 2.58 & 1.32\\
\:\:\:Panel Config. 2 & 8.52 & 3.61 & 2.63 & 1.36\\
\:\:\:Panel Config. 3 & 8.39 & 2.94 & 2.67 & 1.35\\
\cmidrule(l{1em}r{1em}){1-5}
Complex & 4.86 & 0.92 & 2.25 & 1.41\\
\:\:\:w/ Rotated Speaker & 4.33 & 0.83 & 2.13 & 1.41\\
\:\:\:w/ Translated Speaker & 4.38 & 1.19 & 2.22 & 1.44\\
\bottomrule
\end{tabular}
\caption{\name's performance on additional configurations in the \name Dataset, on the task of predicting monaural RIRs and music at an unseen point. Lower is better for all metrics. Errors for RIRs are multiplied by 10. Each \name model is trained on 12 points.}
\label{tab:additionalconfigresults}
\end{table}

%% file: tables/virtualRotation.tex
\begin{table}
\small
\centering
\begin{tabular}{lccccc}
\toprule
 & \multicolumn{2}{c}{RIR} &  \multicolumn{2}{c}{Music} \\
 \cmidrule(lr){2-3} \cmidrule(lr){4-5}
 
\textbf{Room/Configuration} & Mag & ENV & Mag & ENV \\
\midrule

\textbf{Dampened w/ Rotation}\\
\:\:\:Trained on Rot. Data & \textbf{1.14} & \textbf{0.44} & \textbf{1.49} & \textbf{1.36}\\
\:\:\:Trained on Base w/ Virt. Rot. & 1.39 & 0.51 & 1.88 & 1.48 \\
\cmidrule[0.25pt](l{1em}r{1em}){1-5}
\textbf{Hallway w/ Rotation}\\
\:\:\:Trained on Rot. Data & \textbf{8.40} & \textbf{2.86} & \textbf{2.58} & \textbf{1.27}\\
\:\:\:Trained on Base w/ Virt. Rot. & 9.83 & 3.22 & 2.88 & 2.50 \\
\cmidrule[0.25pt](l{1em}r{1em}){1-5}
\textbf{Complex w/ Rotation}\\
\:\:\:Trained on Rot. Data & \textbf{4.33} & \textbf{0.83} & \textbf{2.13} & \textbf{1.41}\\
\:\:\:Trained on Base w/ Virt. Rot. & 4.84 & 0.89 & 2.27 & 1.59 \\

\bottomrule

\end{tabular}
\caption{Results on Virtual Speaker Rotation. Evaluations are done on the test set of the rotated subdataset.}
\label{tab:virtualRot}
\end{table}

%% file: tables/virtualInsertion.tex
\begin{table}
\small
\centering
\begin{tabular}{lccccc}
\toprule
 & \multicolumn{2}{c}{RIR} &  \multicolumn{2}{c}{Music} \\
 \cmidrule(lr){2-3} \cmidrule(lr){4-5}
 
\textbf{Room/Configuration} & Mag & ENV & Mag & ENV \\
\midrule

\multicolumn{5}{l}{\textbf{Damp. $\rightarrow$ Hall. 1.}}\\
\:\:\:Hall. 1 Model & \textbf{8.47} & \textbf{2.99} & \textbf{2.58} & \textbf{1.32}\\
\:\:\:Virtual Insertion & 9.32 & 2.96 & 2.69 & 1.33\\
\multicolumn{5}{l}{\textbf{Damp. $\rightarrow$ Hall. 2.}}\\
\:\:\:Hall. 2 Model & \textbf{8.52} & \textbf{3.61} & \textbf{2.63} & \textbf{1.36}\\
\:\:\:Virtual Insertion & 9.31 & 3.45 & 2.62 & 1.38\\
\multicolumn{5}{l}{\textbf{Hall. 1.$\rightarrow$ Damp.}}\\
\:\:\:Damp. Panel Model & \textbf{1.23} & \textbf{0.600} & \textbf{1.62} & \textbf{1.47}\\
\:\:\:Virtual Insertion & 1.84 & 0.660 & 3.70 & 1.56\\
\multicolumn{5}{l}{\textbf{Hall. 2.$\rightarrow$ Damp.}}\\
\:\:\:Damp. Panel Model & \textbf{1.23} & \textbf{0.600} & \textbf{1.62} & \textbf{1.47}\\
\:\:\:Virtual Insertion & 1.84 & 0.660 & 3.70 & 1.56\\

\bottomrule

\end{tabular}
\caption{Results on Virtual Panel Insertion. `Damp.$\rightarrow$Hall 1.' means that take the \name model from the Hallway Base subdataset (no panel). Then, we virtually insert a panel to simulate the Hallway Panel 1 subdataset, by borrowing the reflection coefficients of the panel from the \name model trained on the Dampened w/ Panel subdataset. We then evaluate the virtual insertion on the recordings from the Hallway Panel 1 subdataset. As a baseline, we compare to a model that is trained on the same panel subdataset that it is tested on.}
\label{tab:virtualInsertion}
\end{table}

%% file: tables/virtualPanelMovement.tex
\begin{table}
\small
\centering
\begin{tabular}{lccccc}
\toprule
 & \multicolumn{2}{c}{RIR} &  \multicolumn{2}{c}{Music} \\
 \cmidrule(lr){2-3} \cmidrule(lr){4-5}
 
\textbf{Room/Configuration} & Mag & ENV & Mag & ENV \\
\midrule

\multicolumn{5}{l}{\textbf{Hall. 1.$\rightarrow$ Hall. 2.}}\\
\:\:\:Baseline & \textbf{8.52} & \textbf{3.61} & \textbf{2.63} & \textbf{1.36}\\
\:\:\:Virtual Panel Relocation & 8.91 & 3.59 & 2.71 & 1.39\\
\multicolumn{5}{l}{\textbf{Hall. 2.$\rightarrow$ Hall 1.}}\\
\:\:\:Baseline & \textbf{8.47} & \textbf{2.99} & \textbf{2.58} & \textbf{1.32}\\
\:\:\:Virtual Panel Relocation & 8.89 & 3.13 & 2.72 & 1.39\\

\bottomrule

\end{tabular}
\caption{Results on Virtual Panel Relocation. `Hall 1.$\rightarrow$ Hall 2.' means that take the \name model from the Hallway Panel 1 subdataset (no panel). Then, we virtually move this panel to its location in the Hallway Panel 2 subdataset. We then evaluate on the recordings from the Hallway Panel 2 subdataset.}
\label{tab:virtualPanelMovement}
\end{table}

%% file: tables/virtualTranslation.tex
\begin{table}
\small
\centering
\begin{tabular}{lccccc}
\toprule
 & \multicolumn{2}{c}{RIR} &  \multicolumn{2}{c}{Music} \\
 \cmidrule(lr){2-3} \cmidrule(lr){4-5}
 
\textbf{Room/Configuration} & Mag & ENV & Mag & ENV \\
\midrule

\textbf{Dampened w/ Translation}\\
\:\:\:Trained on Trans. Data & \textbf{0.68} & \textbf{0.39} & \textbf{0.91} & \textbf{1.18}\\
\:\:\:Trained on Base w/ Virt. Trans. & 1.22 & 0.53 & 1.26 & 1.61 \vspace{2pt}\\
\cmidrule[0.25pt](l{1em}r{1em}){1-5}
\textbf{Hallway w/ Translation}\\
\:\:\:Trained on Trans. Data & \textbf{8.91} & \textbf{3.02} & \textbf{2.84} & \textbf{1.25}\\
\:\:\:Trained on Base w/ Virt. Trans. & 9.28 & 3.05 & \textbf{2.84}    & 1.28 \vspace{2pt}\\
\cmidrule[0.25pt](l{1em}r{1em}){1-5}
\textbf{Complex w/ Translation}\\
\:\:\:Trained on Trans. Data & \textbf{4.38} & \textbf{1.19} & \textbf{2.22} & \textbf{1.44}\\
\:\:\:Trained on Base w/ Virt. Trans. & 4.79 & \textbf{1.19} & 2.24 & 1.54 \vspace{2pt}\\

\bottomrule

\end{tabular}
\caption{Results on Virtual Speaker Translation. Evaluations are done on the test set of the translated subdataset.}
\label{tab:virtualTrans}
\end{table}

%% file: tables/bin.tex
\begin{table*}[!h]
\centering
\resizebox{12.5cm}{!}{
\begin{tabular}{ l ccc ccc ccc ccc }

    \toprule
    
    & \multicolumn{2}{c}{\textbf{Classroom}} & \multicolumn{2}{c}{\textbf{Dampened Room}} & \multicolumn{2}{c}{\textbf{Hallway}} & \multicolumn{2}{c}{\textbf{Complex Room}} \\
    
    \cmidrule(lr){2-3} \cmidrule(lr){4-5} \cmidrule(lr){6-7} \cmidrule(lr){8-9}

    & RIR & Music & RIR & Music & RIR & Music & RIR & Music \\ \midrule 

    NN & 1.27 & 0.516 & 5.64 & 2.57 & 0.062 & 0.034 & 0.345 & 0.166\\
    Linear & 1.29 & 0.531 & 5.48 & 2.09 & \textbf{0.045} & \textbf{0.008} & 0.335 & \textbf{0.157}\\
    DeepIR & 1.10 & 0.529 & 6.20 & 5.90 & 0.048 & 0.036 & 0.350 & 0.397\\
    NAF & 1.93 & 0.743 & 5.93 & 2.37 & 0.108 & 0.012 & 0.320 & 0.176\\
    INRAS & 1.25 & 0.383 & 5.86 & 4.35 & 1.60 & 4.41 & 0.332 & 0.183\\
    \name (ours) & \textbf{0.43} & \textbf{0.091} & \textbf{2.94} & \textbf{0.316} & 0.097 & 0.012 & \textbf{0.287} & 0.288\\
    % OLD Linear & 1.32 & 0.565 & 5.64 & 2.18 & 0.047 & 0.008 & 0.336 & 0.159\\

    \bottomrule
\end{tabular}
}

\caption{
    Experimental results from the task of predicting binaural RIRs and music at an unseen point from a model trained on 12 monoaural RIRs. We use the left-right energy ratio error metric~\cite{chen2023novel}. Lower is better. All errors are multiplied by 10. 
}
\label{tab:binaural_results}
\end{table*}

%% file: tables/more-ablations.tex
\begin{table*}
\centering
\resizebox{17.5cm}{!}{
\begin{tabular}{ l cccc cccc cccc cccc }
    \toprule
    
    & \multicolumn{4}{c}{\textbf{Classroom}} & \multicolumn{4}{c}{\textbf{Dampened Room}} & \multicolumn{4}{c}{\textbf{Hallway}} & \multicolumn{4}{c}{\textbf{Complex Room}} \\ 
    
    \cmidrule(lr){2-5} \cmidrule(lr){6-9} \cmidrule(lr){10-13} \cmidrule(lr){14-17}
    
    & \multicolumn{2}{c}{RIR} & \multicolumn{2}{c}{Music} & \multicolumn{2}{c}{RIR} & \multicolumn{2}{c}{Music} & \multicolumn{2}{c}{RIR} & \multicolumn{2}{c}{Music} & \multicolumn{2}{c}{RIR} & \multicolumn{2}{c}{Music} \\
    
    \cmidrule(lr){2-3} \cmidrule(lr){4-5} \cmidrule(lr){6-7} \cmidrule(lr){8-9} \cmidrule(lr){10-11} \cmidrule(lr){12-13} \cmidrule(lr){14-15} \cmidrule(lr){16-17}
    
    & Mag & ENV & Mag & ENV & Mag & ENV & Mag & ENV & Mag & ENV & Mag & ENV & Mag & ENV & Mag & ENV  \\
    
    \midrule 
    \name & 5.22 & \textbf{0.942} & 2.71 & \textbf{1.36} & \textbf{1.21} & \textbf{0.555} & 1.59 & 1.19 & \textbf{9.13} & 2.95 & 2.59 & 1.25 & \textbf{4.86} & 0.917 & 2.25 & \textbf{1.41}\\
    \:\:\:w/o Time-of-Arrival Perturbation & \textbf{5.19} & 0.962 & \textbf{2.70} & 1.43 & 1.23 & 0.582 & 1.61 & 1.36 & \textbf{9.13} & 2.93 & 2.60 & 1.27 & \textbf{4.86} & \textbf{0.913} & 2.23 & 1.42\\
    \:\:\:w/o Axial Boosting & \textbf{5.19} & 0.969 & 2.71 & 1.43 & 1.22 & \textbf{0.555} & 1.59 & 1.20 & 9.14 & 2.95 & 2.59 & 1.30 & \textbf{4.86} & 0.934 & 2.25 & 1.44\\
    \:\:\:w/o Hop Size 1 Loss & 5.26 & 0.988 & 2.74 & 1.41 & 1.25 & 0.559 & 1.67 & \textbf{1.16} & 9.22 & 2.98 & 2.60 & \textbf{1.24} & 4.90 & 0.962 & 2.27 & 1.42\\
    \:\:\:w/o Interpolation Spline & 5.60 & 0.973 & 2.72 & 1.41 & 1.63 & 0.565 & \textbf{1.53} & \textbf{1.16} & 9.47 & \textbf{2.92} & \textbf{2.56} & \textbf{1.24} & 5.24 & 0.920 & \textbf{2.21} & 1.42\\

    \bottomrule 

\end{tabular}
}
\caption{
    Ablation results from the task of predicting monaural RIRs and music at an unseen point. In the Interpolation Spline ablation, the \emph{Residual Component} and the contributions from explicitly computed reflection paths are simply added together, instead of being blended using the learned temporal spline $\gamma$. Lower is better for all metrics. Errors for RIRs are multiplied by 10.
}
\label{tab:ablations2}
\vspace{-5pt}
\end{table*}

%% file: tables/transmission.tex
\begin{table}
\small
\centering
\begin{tabular}{lccccc}
\toprule
 & \multicolumn{2}{c}{RIR} &  \multicolumn{2}{c}{Music} \\
 \cmidrule(lr){2-3} \cmidrule(lr){4-5}
 
\textbf{Room/Configuration} & Mag & ENV & Mag & ENV \\
\midrule

\textbf{Classroom}\\
\:\:\:\name (ours) & \textbf{5.22} & \textbf{0.942} & \textbf{2.71} & \textbf{1.36}\\
\:\:\:w/ Transmission & 5.23 & 0.951 & 2.72 & \textbf{1.36} \\
\cmidrule[0.25pt](l{1em}r{1em}){1-5}
\textbf{Dampened Room w/ Panel}\\
\:\:\:\name (ours) & \textbf{1.23} & \textbf{0.604} & \textbf{1.62} & 1.47\\
\:\:\:w/ Transmission & \textbf{1.23} & \textbf{0.604} & \textbf{1.62} & \textbf{1.45} \\
\cmidrule[0.25pt](l{1em}r{1em}){1-5}
\textbf{Hallway w/ Panels}\\
\:\:\:\name (ours) & 8.39 & 2.94 & 2.67 & 1.35\\
\:\:\:w/ Transmission & \textbf{8.38} & \textbf{2.92} & \textbf{2.64} & \textbf{1.34} \\ 
\cmidrule[0.25pt](l{1em}r{1em}){1-5}
\textbf{Complex Room}\\
\:\:\:\name (ours) & \textbf{4.86} & 0.917 & 2.25 & 1.41\\
\:\:\:w/ Transmission & \textbf{4.86} & \textbf{0.915} & \textbf{2.24} & \textbf{1.38} \\

\bottomrule

\end{tabular}
\caption{Evaluations of \name vs \name with Transmission modeling. Lower is better for all metrics, and RIR errors are multiplied by 10.}
\label{tab:transmission}
\end{table}

% \begin{table*}[!t]
% \centering
% \resizebox{17.5cm}{!}{
% \begin{tabular}{ l cccc cccc cccc cccc }
%     \toprule
    
%     & \multicolumn{4}{c}{\textbf{Classroom}} & \multicolumn{4}{c}{\textbf{Dampened Room w/ Panel}} & \multicolumn{4}{c}{\textbf{Hallway w/ Panels}} & \multicolumn{4}{c}{\textbf{Complex Room}} \\ 
    
%     \cmidrule(lr){2-5} \cmidrule(lr){6-9} \cmidrule(lr){10-13} \cmidrule(lr){14-17}
    
%     & \multicolumn{2}{c}{RIR} & \multicolumn{2}{c}{Music} & \multicolumn{2}{c}{RIR} & \multicolumn{2}{c}{Music} & \multicolumn{2}{c}{RIR} & \multicolumn{2}{c}{Music} & \multicolumn{2}{c}{RIR} & \multicolumn{2}{c}{Music} \\
    
%     \cmidrule(lr){2-3} \cmidrule(lr){4-5} \cmidrule(lr){6-7} \cmidrule(lr){8-9} \cmidrule(lr){10-11} \cmidrule(lr){12-13} \cmidrule(lr){14-15} \cmidrule(lr){16-17}
    
%     & Mag & ENV & Mag & ENV & Mag & ENV & Mag & ENV & Mag & ENV & Mag & ENV & Mag & ENV & Mag & ENV  \\
    
%     \midrule 
%     \name & \textbf{5.22} & \textbf{0.942} & \textbf{2.71} & \textbf{1.36} & \textbf{1.23} & \textbf{0.604} & \textbf{1.62} & 1.47 & 8.39 & 2.94 & 2.67 & 1.35 & \textbf{4.86} & 0.917 & 2.25 & 1.41\\
%     \:\:\:w/ Transmission & 5.23 & 0.951 & 2.72 & \textbf{1.36} & \textbf{1.23} & \textbf{0.604} & \textbf{1.62} & \textbf{1.45} & \textbf{8.38} & \textbf{2.92} & \textbf{2.64} & \textbf{1.34} & \textbf{4.86} & \textbf{0.915} & \textbf{2.24} & \textbf{1.38}\\
    
%     \bottomrule 

% \end{tabular}
% }
% \caption{
%     Results when modeling surface transmission.
% }
% \label{tab:transmission}
% \vspace{-5pt}
% \end{table*}

%% file: tables/pyroom.tex
\begin{table}
\small
\centering
\begin{tabular}{lccccc}
\toprule
 & \multicolumn{2}{c}{RIR} &  \multicolumn{2}{c}{Music} \\
 \cmidrule(lr){2-3} \cmidrule(lr){4-5}
 
\textbf{Room/Configuration} & Mag & ENV & Mag & ENV \\
\midrule

\textbf{Classroom}\\
\:\:\:\name (ours) & \textbf{5.22} & \textbf{0.942} & \textbf{2.71} & \textbf{1.36}\\
\:\:\:Pyroomacoustics & 18.64 & 3.67 & 3.26 & 1.68\\
\cmidrule[0.25pt](l{1em}r{1em}){1-5}
\textbf{Dampened Room}\\
\:\:\:\name & \textbf{1.21} & \textbf{0.555} & \textbf{1.59} & \textbf{1.19}\\
\:\:\:Pyroomacoustics & 2.14 & 0.798 & 2.17 & 1.96 \\
\cmidrule[0.25pt](l{1em}r{1em}){1-5}
\textbf{Hallway}\\
\:\:\:\name & \textbf{9.13} & \textbf{2.95} & \textbf{2.59} & \textbf{1.25}\\
\:\:\:Pyroomacoustics & 32.01 & 4.03 & 3.39 & 1.70\\

\bottomrule

\end{tabular}
\caption{Comparison of our model against Pyroomacoustics. Lower is better for all metrics, and RIR errors are multiplied by 10.}
\label{tab:pyroom}
\end{table}

%% file: tables/pink.tex
\begin{table*}
\centering
\resizebox{17.5cm}{!}{
\begin{tabular}{ l cccc cccc cccc cccc }
    \toprule
    & \multicolumn{4}{c}{\textbf{Classroom}} & \multicolumn{4}{c}{\textbf{Dampened Room}} & \multicolumn{4}{c}{\textbf{Hallway}} & \multicolumn{4}{c}{\textbf{Complex Room}} \\ \cmidrule(lr){2-5} \cmidrule(lr){6-9} \cmidrule(lr){10-13} \cmidrule(lr){14-17}
& \multicolumn{2}{c}{RIR} & \multicolumn{2}{c}{Music} & \multicolumn{2}{c}{RIR} & \multicolumn{2}{c}{Music} & \multicolumn{2}{c}{RIR} & \multicolumn{2}{c}{Music} & \multicolumn{2}{c}{RIR} & \multicolumn{2}{c}{Music} \\ \cmidrule(lr){2-3} \cmidrule(lr){4-5} \cmidrule(lr){6-7} \cmidrule(lr){8-9} \cmidrule(lr){10-11} \cmidrule(lr){12-13} \cmidrule(lr){14-15} \cmidrule(lr){16-17}
    & Mag & ENV & Mag & ENV & Mag & ENV & Mag & ENV & Mag & ENV & Mag & ENV & Mag & ENV & Mag & ENV \\ \midrule 
    Pink Reg & 5.22 & \textbf{0.942} & \textbf{2.71} & \textbf{1.36} & \textbf{1.21} & \textbf{0.555} & \textbf{1.59} & \textbf{1.19} & \textbf{9.13} & \textbf{2.95} & \textbf{2.59} & \textbf{1.25} & 4.86 & 0.917 & \textbf{2.25} & \textbf{1.41}\\
    No Reg. & 5.22 & 0.973 & 2.76 & 1.47 & 1.23 & 0.579 & 1.62 & 1.33 & 9.17 & 2.99 & 2.71 & 1.35 & \textbf{4.84} & 0.908 & 2.26 & 1.45\\
    Music Reg. & \textbf{5.20} & 0.952 & 2.72 & 1.40 & 1.22 & 0.569 & 1.62 & 1.27 & 9.14 & 2.96 & 2.65 & 1.31 & 4.84 & \textbf{0.903} & 2.25 & 1.42\\
    \bottomrule 

\end{tabular}
}
\caption{
    Comparison of our model trained with no regularization, regularizing by convolving with pink noise, and regularization by convolving with music, on the tasks of monoaural RIR and music prediction. Lower is better for all metrics, and RIR errors are multiplied by 10.
}
\label{tab:pinkconvolve}
\end{table*}

%% file: tables/mic_placement.tex
\begin{table}
\small
\centering
\begin{tabular}{lccccc}
\toprule
 & \multicolumn{2}{c}{RIR} &  \multicolumn{2}{c}{Music} \\
 \cmidrule(lr){2-3} \cmidrule(lr){4-5}
 
\textbf{Training Point Configuration} & Mag & ENV & Mag & ENV \\
\midrule

\textbf{Near} & 5.89 & 1.14 & 3.25 & 1.61\\
\textbf{Spread} & \textbf{5.39} & \textbf{0.976} & \textbf{2.80} & \textbf{1.36}\\
\textbf{Corner} & 5.88 & 1.07 & 3.12 & 1.41\\

\bottomrule

\end{tabular}
\caption{Evaluations of \name on different datasets of size 6, with varying spatial distributions. All microphone locations are selected from $Z=0.98$, and all locations used for testing and evaluation are also selected from $Z=0.98$. Lower is better for all metrics, and RIR errors are multiplied by 10.}
\label{tab:mic_placement}
\end{table}